\documentclass[secthm]{elsart}
\usepackage{latexsym}
\usepackage{url}
\usepackage{amssymb}
\usepackage{gastex}

\newtheorem{pro-example}[thm]{Example}
\newenvironment{expl}{\begin{pro-example}\rm}{\cqfd\end{pro-example}}

\newtheorem{pro-remark}[thm]{Remark}
\newenvironment{remark}{\begin{pro-remark}\rm}{\cqfd\end{pro-remark}}

\newenvironment{Proof}{\rm \trivlist \item[\hskip \labelsep{\bf
Proof.}]}{\cqfd\endtrivlist}

\def\cqfd{\skip10=\parfillskip\parfillskip=0pt
\enspace\hfill\symbolecqfd\par\parfillskip=\skip10\par\medskip}
\def\symbolecqfd{\rlap{$\sqcap$}$\sqcup$}
\def\preuve{\begin{Proof}}
\def\proof{\begin{Proof}}
\def\eop{\end{Proof}}

\def\pgs{$pg$-pairs}
\def\0{{\bf 0}}
\def\1{{\bf 1}}

\def\bW{{\bf W}}
\def\cW{{\mathcal{W}}}

\def\cV{\mathcal{V}}

\def\bV{{\bf V}}
\def\n{{\bf n}}

\def\k{{\bf k}}

\def\K{{\bf K}}

\def\L{{\cal L}}

\def\Suc{{\bf Succ }}
\def\false{{\sf false}}
\def\true{{\sf true}}

\def\Box{\hbox{\rlap{$\sqcap$}$\sqcup$}}

\def\inv{^{-1}}
\let\phi\varphi

\let\ol\overline

\def\sfor{{\sf or}}

\def\note#1{}

\def\id{\textsf{id}}
\def\pg{\textsf{pg}}
\def\var{{\sf var}}
\def\psv{{\sf psv}}

\def\val{{\sf val}}

\def\EX{\textsf{EX}}
\def\EF{\textsf{EF}}
\def\path{{\sf path}}

\def\next{{\sf next}}

\def\p{{\bf p}}

\def\I{{\bf I}}
\def\H{{\bf H}}
\def\S{{\bf S}}
\def\P{{\bf P}}
\def\L{{\bf L}}
\def\U{{\bf U}}
\def\HSP{{\bf HSP}}
\def\ULHSP{{\bf ULHSP}}

\def\B{\mathbb{B}}

\def\ellbf{{\textbf{l}}}

\begin{document}

% 
% \centerline{\LARGE Algebraic recognizability of regular tree
% languages\protect\footnote{The first author acknowledges partial
% support from the National Foundation of Hungary for Scientific
% Research, grant T46686. Part of this work was done while the second
% author was an invited professor at the University of Nebraska in
% Lincoln. The same author also acknowledges partial support from the
% \textit{ACI S\'ecurit\'e Informatique} (projet \textsc{versydis}) of
% the French Minist\`ere de la Recherche.}}
% 
% \vskip .5cm
% 
% \centerline{\Large Zolt\'an \'Esik\protect\footnote{%
% Department of Computer Science, University of Szeged, Hungary.
% \url{ze@inf.u-szeged.hu}}\ \protect\footnote{%
% Research Group on Mathematical Linguistics, Rovira i Virgili
% University, Tarragona, Spain.}
% % 
% \qquad
% % 
% Pascal Weil\protect\footnote{%
% LaBRI - CNRS, 351 cours de la Lib\'eration, 33405 Talence Cedex,
% France. \url{pascal.weil@labri.fr}}}
% 
% 

\begin{frontmatter}
    
\title{Algebraic recognizability of regular tree languages}

\author[Szeged]{Zolt\'an \'Esik\thanksref{ZE}}
\thanks[ZE]{Partial support from the National Foundation of Hungary for
Scientific Research, grant T46686 is gratefully acknowledged.}
\ead{ze@inf.u-szeged.hu}
% 
% \address[Szeged]{Research Group on Mathematical Linguistics, Rovira i Virgili
% University, Tarragona, Spain.}
% 
% 
\author[LaBRI]{Pascal Weil\thanksref{PW}}
\address[Szeged]{Department of Computer Science, University of Szeged,
Hungary\\
Research Group on Mathematical Linguistics, Rovira i Virgili
University, Tarragona, Spain.}
\address[LaBRI]{LaBRI, CNRS, Universit\'e Bordeaux-1, France}
\ead{pascal.weil@labri.fr}
\thanks[PW]{Partial support from the \textit{ACI S\'ecurit\'e Informatique}
(projet \textsc{versydis}) of the French Minist\`ere de la Recherche is
gratefully acknowledged. Part of this work was done while P. Weil was
an invited professor at the University of Nebraska in Lincoln.}

\vskip 1cm

\begin{abstract}
    We propose a new algebraic framework to discuss and classify
    recognizable tree languages, and to characterize interesting
    classes of such languages. Our algebraic tool, called preclones,
    encompasses the classical notion of syntactic $\Sigma$-algebra or
    minimal tree automaton, but adds new expressivity to it. The main
    result in this paper is a variety theorem \textit{\`a la}
    Eilenberg, but we also discuss important examples of logically
    defined classes of recognizable tree languages, whose
    characterization and decidability was established in recent papers
    (by Benedikt and S\'egoufin, and by Boja\'nczyk and Walukiewicz)
    and can be naturally formulated in terms of pseudovarieties of
    preclones. Finally, this paper constitutes the foundation for
    another paper by the same authors, where first-order definable tree
    languages receive an algebraic characterization.
\end{abstract} 

\end{frontmatter}

\section{Introduction}

The notion of recognizability emerged in the 1960s (Eilenberg, Mezei,
Wright, and others, cf. \cite{EilenbergWright,MezeiWright}) and has
been the subject of considerable attention since, notably because of
its close connections with automata-theoretic formalisms and with
logical definability, cf. \cite{Buchi,Doner,Elgot,ThatcherWright} for
some early papers.

Recognizability was first considered for sets (languages) of finite
words, cf. \cite{Eilenberg} and the references contained in op. cit.
The general idea is to use the algebraic structure of the domain, say,
the monoid structure on the set of all finite words, to describe some
of its subsets, and to use algebraic considerations to discuss the
combinatorial or logical properties of these subsets. More precisely, a
set of words is said to be recognizable if it is a union of classes in
a (locally) finite congruence. The same concept was adapted to the case
of finite trees, traces, finite graphs, etc, cf.
\cite{EilenbergWright,MezeiWright,Diekert,Courcelle}, where it always
entertains close connections with logical definability
\cite{CourcelleHdbook,CourcelleWeil}.

It follows rather directly from this definition of (algebraic)
recognizability that a finite -- or finitary -- algebraic structure can
be canonically associated with each recognizable subset $L$, called its
syntactic structure. Moreover, the algebraic properties of the
syntactic structure of $L$ reflect its combinatorial and logical
properties. The archetypal example is that of star-free languages of
finite words: they are exactly the languages whose syntactic monoid is
aperiodic, cf. \cite{Schutzenberger}. They are also exactly the
languages that can be defined by a first-order sentence of the
predicate $<$ ($FO[<]$), cf. \cite{McNaughtonPapert}, and the languages
that can be defined by a temporal logic formula, cf.
\cite{Kamp,Gabbayetal,Cohenetal}. In particular, every algorithm we
know for deciding the $FO[<]$-definability of a regular language $L$,
works by checking, more or less explicitly, whether the syntactic
monoid of $L$ is aperiodic.

Let $\Sigma$ be a ranked alphabet. In this paper, we are interested in
sets of finite $\Sigma$-labeled trees, or tree languages. It has been
known since the 1960s \cite{EilenbergWright,MezeiWright,Doner} that the
tree languages that are definable in monadic second order logic are
exactly the so-called regular tree languages, that is, those accepted
by bottom-up tree automata. Moreover, deterministic tree automata
suffice to accept these languages, and each regular tree language
admits a unique, minimal deterministic automaton. From the algebraic
point of view, the set of all $\Sigma$-labeled trees can be viewed in a
natural way as a (free) $\Sigma$-algebra, where $\Sigma$ is now seen as
a signature. Moreover, a deterministic bottom-up tree automaton can be
identified with a finite $\Sigma$-algebra, with some distinguished
(final) elements. Thus regular tree languages are also the recognizable
subsets of the free $\Sigma$-algebra.

The situation however is not entirely satisfying, because we know very
little about the structure of finite $\Sigma$-algebras, and very few
classes of tree languages have been characterized in algebraic terms,
see \cite{Heuter,Potthoff1,Potthoff95} for attempts to use
$\Sigma$-algebra-theoretic considerations (and some variants) for the
purpose of classifying tree languages. In particular, the important
problem of deciding whether a regular tree language is $FO[<]$-definable
remained open \cite{Potthoff95}. Based on the word language case, it is
tempting to guess that an answer to this problem ought to be found
using algebraic methods.

In this paper, we introduce a new algebraic framework to handle tree
languages. More precisely, we consider algebras called preclones (they
lack some of the operations and axioms of clones \cite{Deneckeetal}).
Precise definitions are given in Section~\ref{sec-preclones}. Let us
simply say here that, in contrast with the more classical monoids or
$\Sigma$-algebras, preclones have infinitely many sorts, one for each
integer $n\ge 0$. As a result, there is no nontrivial finite preclone.
The corresponding notion is that of finitary preclones, that have a
finite number of elements of each sort. An important class of preclones
is given by the transformations $T(Q)$ of a set $Q$. The elements of
sort (or rank) $n$ are the mappings from $Q^n$ into $Q$ and the
(preclone) composition operation is the usual composition of mappings.
Note that $T(Q)$ is finitary if $Q$ is finite.

It turns out that the finite $\Sigma$-labeled trees can be identified
with the $0$-sort of the free preclone generated by $\Sigma$. The
naturally defined syntactic preclone of a tree language $L$ is finitary
if and only if $L$ is regular. In fact, if $S$ is the syntactic
$\Sigma$-algebra of $L$, the syntactic preclone is the sub-preclone of
$T(S)$ generated by the elements of $\Sigma$ (if $\sigma\in\Sigma$ is
an operation of rank $r$, it defines a mapping from $S^r$ into $S$, and
hence an element of sort $r$ in $T(S)$). Note that this provides an
effectively constructible description of the syntactic preclone of $L$.

It is important to note that the class of recognizable tree languages
in the preclone-theoretic sense, is exactly the same as the usual one
-- we are simply adding more algebraic structure to the finitary
minimal object associated with a regular tree language, and thus, we
give ourselves a more expressive language to capture families of tree
languages.

In order to justify the introduction of such an algebraic framework, we
must show not only that it offers a well-structured framework, that
accounts for the basic notions concerning tree languages, but also that
it allows the characterization of interesting classes of tree languages.
The first objective is captured in the definition of varieties of tree
languages, and their connection with pseudovarieties of finitary
preclones, by means of an Eilenberg-type theorem. This is not
unexpected, but it requires combinatorially much more complex proofs
than in the classical word case, the details of which can be found
below in Section~\ref{sec eilenberg}.

As for the second objective, we offer several elements. First the
readers will find in this paper a few simple but hopefully illuminating
examples, which illustrate similarities and differences with the
classical examples from the theory of word languages. Second, we
discuss a couple of important recent results on the characterization of
certain classes of tree languages: one concerns the tree languages that
are definable in the first-order logic of successors ($FO(\Suc)$), and
is due to Benedikt and S\'egoufin \cite{BenediktSegoufin}; the second
one concerns the tree languages defined in the logics $\EF$ and $\EX$,
and is due to Boja\'nczyk and Walukiewicz \cite{BW}. Neither of these
remarkable results can be expressed directly in terms of syntactic
$\Sigma$-algebras; neither mentions preclones (of course) but both use
mappings of arity greater than 1 on $\Sigma$-algebras, that is, they
can be naturally expressed in terms of preclones, as we explain in
Sections~\ref{sec FOSuc} and \ref{sec EFEX}. It is also very
interesting to note that the conditions that characterize these various
classes of tree languages include the semigroup-theoretic
characterization of their word language analogues, but cannot be
reduced to them.

Another such result, and that was our original motivation to
introduce the formalism of preclones, is a nice algebraic
characterization of $FO[<]$-definable tree languages (and a number of
extensions of $FO[<]$, such as the introduction of additional, modular
quantifiers), briefly discussed in Section~\ref{sec FOinf}. Let us say
immediately that we do not know yet whether this characterization can
be turned into a decision algorithm! In order to keep this paper within
a reasonable number of pages, this characterization will be the subject
of another paper by the same authors \cite{EWpreparation}. The main
results of this upcoming paper can be found, along with an outline of
the present paper, in \cite{FSTTCS}.

To summarize the plan of the paper, Section~\ref{algebraic} introduces
the algebraic framework of preclones, discussing in particular the
all-important cases of free preclones, in which tree languages live
(Section~\ref{sec free preclone}), and of preclones associated with
tree automata (Section~\ref{sec examples preclones}). Section~\ref{sec
representation} discusses in some details the notion of finite
determination for a preclone, a finiteness condition different from
being finitary, which is crucial in the sequel. Section~\ref{sec
magmoids} is included for completeness (and can be skipped at
first reading): its aim is to make explicit the connection between our
preclones and other known algebraic structures, namely magmoids and
strict monoidal categories.

Recognizable tree languages are the subject of Section~\ref{grand sec
recog}. Here tree languages are meant to be any subset of some $\Sigma
M_{k}$, and the preclone structure on $\Sigma M$ naturally induces a
notion of recognizability, as well as a notion of syntactic preclone
(Section~\ref{sec recog}). As pointed out earlier, the usual
recognizable tree languages, that is, subsets of $\Sigma M_{0}$, fall
nicely in this framework, and there is a tight connection between the
minimal automaton of such a language and its syntactic preclone
(Section~\ref{sec regular tree}). Specific examples are given in
Section~\ref{sec more examples trees}.

Pseudovarieties of finitary preclones are discussed in detail in
Section~\ref{sec psv}. As it turns out, this notion is not a direct
translate of the classical notion for semigroups or monoids, due to the
infinite number of sorts. The technical treatment of these classes is
rather complex, and we deal with it thoroughly, since it is the
foundation of our construction. We show in particular that
pseudovarieties are characterized by their finitely determined elements
(Section~\ref{sec psv vs fd}), and we describe the pseudovarieties
generated by a given set of finitary preclones, showing in particular
that membership in a 1-generated pseudovariety is decidable
(Section~\ref{psv generated}).

Finally, we introduce varieties of tree languages and we establish the
variety theorem in Section~\ref{sec eilenberg}. Section~\ref{sec
examples tree varieties} presents the examples described above, based
on the results by Benedikt and S\'egoufin \cite{BenediktSegoufin} and
by Boja\'nczyk and Walukiewicz \cite{BW}.

%%%%%%%%%%%%%%%%%%%%%%%%%%%%%%%%%
\section{The algebraic framework}\label{algebraic}

In this section, we introduce the notion of preclones, a multi-sorted
kind of algebra which is our central tool in this paper.
In the sequel, if $n$ is an integer, $[n]$ denotes the set of integers
$\{1,\ldots,n\}$. In particular, $[0]$ denotes the empty set.

%%%%%%%%%%%%%%%%%%%%%%%%%%%%%%%%%
\subsection{Preclones and preclone-generators pairs}\label{sec-preclones}

Let $Q$ be a set and let $T_{n}(Q)$ denote the set of $n$-ary
transformations of $Q$, that is, mappings from $Q^n$ to $Q$. Let then
$T(Q)$ be the sequence of sets of transformations $T(Q) =
(T_{n}(Q))_{n\ge 0}$, which will be called the \textit{preclone of
transformations} of $Q$. The set $T_{1}(Q)$ of transformations of $Q$
is a monoid under the composition of functions. Composition can be
considered on $T(Q)$ in general: if $f\in T_{n}(Q)$ and $g_{i}\in
T_{m_{i}}(Q)$ ($1\leq i\leq n$), then the composite $h =
f(g_{1},\ldots,g_{n})$, defined in the natural way, is an element of
$T_{m}(Q)$ where $m = \sum_{i\in[n]}m_{i}$:
\begin{eqnarray*}
h(q_{1,1},\ldots,q_{n,m_n}) &=& 
f(g_1(q_{1,1},\ldots,q_{1,m_1}),\ldots, g_n(q_{n,1},\ldots,q_{n,m_n}))\\
\end{eqnarray*}
for all $q_{i,j} \in Q$, $1 \leq i \leq n$, $1 \leq j \leq m_i$. This
composition operation and its associativity properties are exactly what
is captured in the notion of a preclone.

In general, a \textit{preclone} is a many-sorted algebra $S
=((S_n)_{n\ge 0},\bullet,\1)$. The elements of the sets $S_{n}$, where
$n$ ranges over the nonnegative integers, are said to have
\textit{rank} $n$. The \textit{composition operation} $\bullet$
associates with each $f\in S_n$ and $g_1\in S_{m_1},\ldots,g_n\in
S_{m_n}$, an element $\bullet(f,g_1,\ldots,g_n)\in S_m$, of rank
$m=\sum_{i \in [n]}m_i$. We usually write $f\cdot (g_1\oplus \cdots
\oplus g_n)$ for $\bullet(f,g_1,\ldots,g_n)$. Finally, the constant
$\1$ is in $S_1$. Moreover, we require the following three equational
axioms:
\begin{eqnarray}
(f\cdot (g_1\oplus\cdots \oplus g_n))\cdot (h_1\oplus \cdots\oplus h_m)
&=&
f\cdot((g_1\cdot \ol{h}_1)\oplus \cdots \oplus (g_n\cdot
\ol{h}_n)),\hbox{\qquad}
\end{eqnarray}
where $f,g_1,\ldots,g_n$ are as above, $h_j \in S_{k_j}$ ($j \in [m]$),
and if we denote $\sum_{j\in[i]}m_{j}$ by $m_{[i]}$, then
$\ol{h}_i=h_{m_{[i-1]}+1} \oplus \cdots \oplus h_{m_{[i]}}$ for each $i \in
[n]$;
\begin{eqnarray}
\1\cdot f &=& f\\
f \cdot (\1 \oplus \cdots \oplus \1) &=& f,
\end{eqnarray}
where $f\in S_n$ and $\1$ appears $n$ times on the left hand side of
the last equation.

Note that Axiom (1) generalizes associativity, and Axioms (2) and (3)
can be said to state that $\1$ is a neutral element.

\begin{remark}
    The elements of rank 1 of a preclone form a monoid.
\end{remark}    

It is immediately verified that $T(Q)$, the preclone of transformation
of a set $Q$, is indeed a preclone for the natural composition of
functions, with the identity function $\id_{Q}$ as $\1$. Preclones are
an abstraction of sets of $n$-ary transformations of a set, which
generalizes the abstraction from transformation monoids to monoids.

\begin{remark}
    Clones \cite{Deneckeetal}, or equivalently, Lawvere theories
    \cite{BEbook,Esik} are another more classical abstraction. Readers
    interested in the comparison between clones and preclones will have
    no difficulty tracing their differences in the sequel. We will
    simply point out here the fact that, in contrast with the
    definition of the clone of transformations of $Q$, each of the $m$
    arguments of the composite $f(g_{1},\ldots,g_{n})$ above is
    used in exactly one of the $g_{i}$'s, the first $m_{1}$ in $g_{1}$,
    the next $m_{2}$ in $g_{2}$, etc.
\end{remark}

We observe that a preclone with at least one element of rank greater 
than 1 must have elements of arbitrarily high rank, and hence cannot 
be finite. We say that a preclone $S$ is \emph{finitary} if and only 
if each $S_{n}$ is finite. For instance, the preclone of
transformations $T(Q)$ is finitary if and only if the set $Q$ is
finite.

The notions of \emph{morphism} between preclones, \emph{sub-preclone},
\emph{congruence} and \emph{quotient} are defined as usual
\cite{Graetzer,Wechler}. Note that, as is customary for multi-sorted
algebras, a morphism maps elements of rank $n$ to elements of the same
rank, and a congruence only relates elements of the same rank.

To facilitate discussions, we introduce the following short-hand
notation. An $n$-tuple $(g_1,\ldots,g_n)$ of elements of $S$ will often
be written as a formal $\oplus$-sum: $g_1 \oplus \cdots \oplus g_n$.
Moreover, if $g_{i}\in S_{m_{i}}$ ($1\leq i\leq n$), we say that
$g_{1}\oplus\cdots\oplus g_{n}$ has \emph{total rank} $m =
\sum_{i\in[n]}m_{i}$. Finally, we denote by $S_{n,m}$ the set of all
$n$-tuples of total rank $m$. With this notation, $S_{1,n} = S_{n}$.
The $n$-tuple $\1\oplus\cdots\oplus\1\in S_{n,n}$ is denoted by $\n$.
If $G$ is a subset of $S$, we also denote by $G_{n,m}$ the set of
$n$-tuples of elements of $G$, of total rank $m$.

Observe that a preclone morphism $\phi\colon S\rightarrow T$, naturally
extends to a map $\phi\colon S_{n,m}\to T_{n,m}$ for each $n,m\ge 0$,
by mapping $g_{1}\oplus\cdots\oplus g_{n}$ to
$\phi(g_{1})\oplus\cdots\oplus\phi(g_{n})$.

For technical reasons, it will often be preferable to work with pairs
$(S,A)$ consisting of a preclone $S$ and a (possibly empty) set $A$ of
generators of $S$. We call such pairs \emph{preclone-generators pairs},
or \pgs. The notions of morphisms and congruences must be revised
accordingly: in particular, a morphism of \pgs\ from $(S,A)$ to $(T,B)$
must map $A$ into $B$. A $pg$-pair $(S,A)$ is said to be finitary if $S$ is
finitary and $A$ is finite.

Besides preclones of transformations of the form $T_{n}(Q)$,
fundamental examples of preclones and \pgs\ are the free preclones and
the preclones associated with a tree automaton. These are discussed in
the next sections.

%%%%%%%%%%%%%%%%%%%%%%
\subsection{Trees and free preclones}\label{sec free preclone}

Let $\Sigma$ be a ranked alphabet, say, $\Sigma = (\Sigma_{n})_{n\geq
0}$, and let $(v_{k})_{k\ge 1}$ be a sequence of variable names. We let
$\Sigma M_{n}$ be the set of finite trees whose inner nodes are labeled
by elements of $\Sigma$ (according to their rank), whose leaves are
labeled by elements of $\Sigma_{0}\cup\{v_{1},\ldots,v_{n}\}$, and
whose \textit{frontier} (the left to right sequence of variables
appearing in the tree) is the word $v_{1}\cdots v_{n}$: that is, each
variable occurs exactly once, and in the natural order. Note that
$\Sigma M_{0}$ is the set of finite $\Sigma$-labeled trees. We let
$\Sigma M = (\Sigma M_{n})_{n}$.

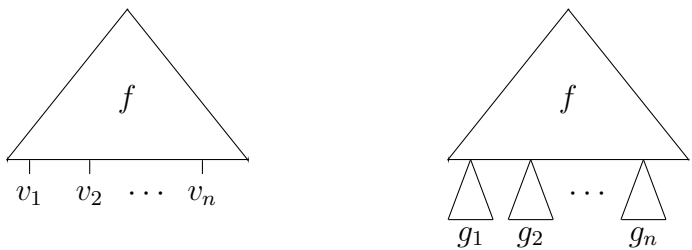
\begin{picture}(49,38)(0,-38)
% \put(0,-45){\framebox(49,45){}}
\node[Nw=0.1,Nh=0.0,Nmr=0.0](n0)(24.0,-5.0){}

\node[Nw=0.0,Nh=0.0,Nmr=0.0](n1)(8.0,-25.0){}

\node[Nw=0.0,Nh=0.0,Nmr=0.0](n2)(40.0,-25.0){}

\drawedge[AHnb=0](n0,n1){ }

\drawedge[AHnb=0](n1,n2){ }

\drawedge[AHnb=0](n0,n2){ }

\node[Nframe=n,Nw=0.0,Nh=0.0,Nmr=0.0](n3)(11.0,-25.0){}

\node[Nframe=n,ExtNL=y,NLangle=270.0,NLdist=1.5,Nw=0.2,Nh=0.2,Nmr=0.0](n4)(11.0,-27.0){$v_1$}

\node[Nframe=n,Nw=0.0,Nh=0.0,Nmr=0.0](n5)(19.0,-25.0){}

\node[Nframe=n,ExtNL=y,NLangle=270.0,NLdist=1.5,Nw=0.2,Nh=0.2,Nmr=0.0](n6)(19.0,-27.0){$v_2$}

\node[Nframe=n,ExtNL=y,NLangle=270.0,NLdist=1.5,Nw=0.2,Nh=0.2,Nmr=0.0](n9)(26.5,-27.0){$\cdots$}

\node[Nframe=n,Nw=0.0,Nh=0.0,Nmr=0.0](n7)(34.0,-25.0){}

\node[Nframe=n,ExtNL=y,NLangle=270.0,NLdist=1.5,Nw=0.2,Nh=0.2,Nmr=0.0](n8)(34.0,-27.0){$v_n$}

\drawedge[AHnb=0](n3,n4){ }

\drawedge[AHnb=0](n5,n6){ }

\drawedge[AHnb=0](n7,n8){ }

\node[Nframe=n,NLangle=0.0](n9)(24.0,-17.0){$f$}

\end{picture}
\qquad
\begin{picture}(49,38)(0,-38)
% \put(0,-45){\framebox(49,45){}}
\node[Nw=0.1,Nh=0.0,Nmr=0.0](n0)(24.0,-5.0){}

\node[Nw=0.0,Nh=0.0,Nmr=0.0](n1)(8.0,-25.0){}

\node[Nw=0.0,Nh=0.0,Nmr=0.0](n2)(40.0,-25.0){}

\drawedge[AHnb=0](n0,n1){ }

\drawedge[AHnb=0](n1,n2){ }

\drawedge[AHnb=0](n0,n2){ }

\node[Nframe=n,Nw=0.0,Nh=0.0,Nmr=0.0](n3)(11.0,-25.0){}

% \node[Nframe=n,ExtNL=y,NLangle=270.0,NLdist=1.5,Nw=0.2,Nh=0.2,Nmr=0.0](n4)(12.0,-34.0){$x_1$}

\node[Nframe=n,Nw=0.0,Nh=0.0,Nmr=0.0](n5)(19.0,-25.0){}

% \node[Nframe=n,ExtNL=y,NLangle=270.0,NLdist=1.5,Nw=0.2,Nh=0.2,Nmr=0.0](n6)(19.0,-34.0){$x_2$}

\node[Nframe=n,ExtNL=y,NLangle=270.0,NLdist=1.5,Nw=0.2,Nh=0.2,Nmr=0.0](n9)(26.5,-27.0){$\cdots$}

\node[Nframe=n,Nw=0.0,Nh=0.0,Nmr=0.0](n7)(34.0,-25.0){}

% \node[Nframe=n,ExtNL=y,NLangle=270.0,NLdist=1.5,Nw=0.2,Nh=0.2,Nmr=0.0](n8)(34.0,-34.0){$x_n$}

% \drawedge[AHnb=0](n3,n4){ }

% \drawedge[AHnb=0](n5,n6){ }

% \drawedge[AHnb=0](n7,n8){ }

\node[Nframe=n,NLangle=0.0](n9)(24.0,-17.0){$f$}

\node[Nframe=n,Nw=0.0,Nh=0.0,Nmr=0.0](n10)(8.0,-33.0){}

\node[Nframe=n,Nw=0.0,Nh=0.0,Nmr=0.0](n11)(14.0,-33.0){}

\node[Nframe=n,Nw=0.0,Nh=0.0,Nmr=0.0](n12)(16.0,-33.0){}

\node[Nframe=n,Nw=0.0,Nh=0.0,Nmr=0.0](n13)(22.0,-33.0){}

\node[Nframe=n,Nw=0.0,Nh=0.0,Nmr=0.0](n14)(31.0,-33.0){}

\node[Nframe=n,Nw=0.0,Nh=0.0,Nmr=0.0](n15)(37.0,-33.0){}

\drawedge[AHnb=0](n3,n10){}
\drawedge[AHnb=0](n3,n11){}
\drawedge[AHnb=0](n11,n10){$g_{1}$}

\drawedge[AHnb=0](n5,n12){}
\drawedge[AHnb=0](n5,n13){}
\drawedge[AHnb=0](n13,n12){$g_{2}$}

\drawedge[AHnb=0](n7,n14){}
\drawedge[AHnb=0](n7,n15){}
\drawedge[AHnb=0](n15,n14){$g_{n}$}

\end{picture}

If $f\in \Sigma M_{n}$ and $g_{1},\ldots,g_{n}\in \Sigma M$, the
composite tree $f\cdot (g_{1}\oplus\cdots\oplus g_{n})$ is obtained by
substituting the root of the tree $g_{i}$ for the variable $v_{i}$ in
$f$ for each $i$, and renumbering consecutively the variables in the
frontiers of $g_{1},\ldots, g_{n}$. Let also $\1\in\Sigma M_{1}$ be the
tree with a single vertex, labeled $v_{1}$. Then $(\Sigma M,\cdot,\1)$
is a preclone.

Each letter $\sigma\in\Sigma$ of rank $n$ can be identified with the
tree with root labeled $\sigma$, where the root's children are leaves
labeled $v_{1},\ldots,v_{n}$. It is easily verified that every
rank-preserving map from $\Sigma$ to a preclone $S$ can be extended in
a unique fashion to a preclone morphism from $\Sigma M$ into $S$. That
is:

\begin{prop}
      $\Sigma M$ is the free preclone generated by $\Sigma$,
and $(\Sigma M, \Sigma)$ is the free $pg$-pair generated by $\Sigma$.
\end{prop}

\begin{remark}
    If $\Sigma_{n} = \emptyset$ for each $n\ne 1$, then $\Sigma M_{n} =
    \emptyset$ for all $n\ne 1$, and $\Sigma M_{1}$ can be assimilated
    with the set of all finite words on the alphabet $\Sigma_{1}$.
    
    If  at least one $\Sigma_n$ with $n>1$ is nonempty, then
    infinitely many $\Sigma M_{n}$ are nonempty, and if in addition
    $\Sigma_{0}\ne\emptyset$, then each $\Sigma M_n$ is nonempty.
\end{remark}

%%%%%%%%%%%%%%%%%%%%%%
\subsection{Examples of preclones}\label{sec examples preclones}

We already discussed preclones of transformations and free preclones.
The next important class of examples is that of preclones (and \pgs) 
associated with $\Sigma$-algebras and tree automata. We also discuss 
a few simple examples of preclones that will be useful in the sequel.

%%%%%%%%%%%%%%%%%%%%%%
\subsubsection{Preclone associated with an automaton}\label{preclone vs
tree automata}

Let $\Sigma$ be a ranked alphabet as above and let $Q$ be a
\emph{$\Sigma$-algebra}: that is, $Q$ is a set and each element
$\sigma\in \Sigma_{n}$ defines an $n$-ary transformation of $Q$,
\textit{i.e.}, a mapping $\sigma^{Q}\colon Q^n\rightarrow Q$. Recall
that $Q$, equipped with a set $F \subseteq Q$ of \emph{final states},
can also be described as a (deterministic, bottom-up) \emph{tree
automaton} accepting trees in $\Sigma M_0$, cf.
\cite{Doner,ThatcherWright,GecsegSteinby,GSHB,tata}.

More precisely, the mapping $\sigma\mapsto \sigma^Q$ induces a morphism
of $\Sigma$-algebras from $\Sigma M_{0}$, viewed here as the initial
$\Sigma$-algebra (\textit{i.e.}, the algebra of $\Sigma$-terms), to
$Q$, say, $\val\colon \Sigma M_{0}\to Q$, and the tree language
accepted by $Q$ is the set $\val\inv(F)$ of trees which evaluate to an
element of $F$.
    
Now, since the elements of $\Sigma_{n}$ can be viewed also as elements
of $T_{n}(Q)$, the mapping $\sigma\mapsto\sigma^Q$ also extends to a
preclone morphism $\tau\colon\Sigma M\rightarrow T(Q)$, whose
restriction to the rank 0 elements is exactly the morphism $\val$.
The range of $\tau$ is called the \textit{preclone associated with
$Q$}, and the \textit{$pg$-pair associated with $Q$}, written $\pg(Q)$,
is the pair $(\tau(\Sigma M), \tau(\Sigma))$.

We observe in particular that a morphism of $\Sigma$-algebras
$\phi\colon Q\rightarrow Q'$ induces a morphism of \pgs\ $\phi\colon
\pg(Q)\rightarrow \pg(Q')$ in a functorial way.

Conversely, if $Q$ is a set and $\tau\colon\Sigma M\rightarrow T(Q)$ is
a preclone morphism such that $\tau(\Sigma M_{0}) = Q$, letting
$\sigma^Q = \tau(\sigma)$ endows the set $Q$ with a structure of
$\Sigma$-algebra, for which the associated preclone is the range of
$\tau$.

In the sequel, when discussing decidability issues concerning
preclones, we will say that a preclone is \emph{effectively given} if
it is given as the preclone associated with a finite $\Sigma$-algebra
$Q$, that is, by a finite set of generators in $T(Q)$. By
definition, such a preclone is finitary.
    
%%%%%%%%%%%%%%%%%%%%%%
\subsubsection{Simple examples of preclones}\label{simple examples}

The following examples of preclones and \pgs\ will be discussed
throughout the rest of this paper.

\begin{expl}\label{define Texists}
    Let $\B$ be the 2-element set $\B = \{\true,\false\}$, and let
    $T_{\exists}$ be the subset of $T(\B)$ whose rank $n$ elements are
    the $n$-ary \sfor\ function and the $n$-ary constant $\true$,
    written respectively $\sfor_{n}$ and $\true_{n}$. One verifies
    easily that $T_{\exists}$ is a preclone, which is generated by the
    binary $\sfor_{2}$ function and the nullary constants $\true_{0}$
    and $\false_{0}$. That is, if $\Sigma$ consists of these three
    generators, $T_{\exists}$ is the preclone associated with the
    $\Sigma$-automaton whose state set is $\B$.
    
    Moreover, the rank 1 elements of $T_{\exists}$ form a 2-element
    monoid, isomorphic to the multiplicative monoid $\{0,1\}$, and
    known as $U_{1}$ in the literature on monoid theory, e.g.
    \cite{Pin}.
\end{expl}    

\begin{expl}\label{define Tp}
    Let $p$ be an integer, $p\geq 2$ and let $\B_{p} =
    \{0,1,\ldots,p-1\}$. We let $T_{p}$ be the subset of $T(\B_{p})$
    whose rank $n$ elements ($n\geq 0$) are the mappings
    $f_{n,r}\colon(r_{1},\ldots,r_{n})\mapsto r_{1}+\cdots+r_{n}+r
    \bmod p$ for $0\leq r < p$. It is not difficult to verify that
    $T_{p}$ is a preclone, and that it is generated by the nullary
    constant 0, the unary increment function $f_{1,1}$ and the binary
    sum $f_{2,0}$.

    As in Example~\ref{define Texists}, $T_{p}$ can be seen as the
    preclone associated with a $p$-state automaton. Moreover, its rank
    1 elements form a monoid isomorphic to the cyclic group of order
    $p$.
\end{expl}    

\begin{expl}\label{define Tpath}
    Let again $\B = \{\true,\false\}$, and let $T_{\path}$ be the
    subset of $T(\B)$ whose rank $0$ elements are the nullary constants
    $\true_{0}$ and $\false_{0}$, and the rank $n$ elements ($n > 0$)
    are the $n$-ary constants $\true_{n}$ and $\false_{n}$, and the
    $n$-ary partial disjunctions $\sfor_{P}$ (if $P\subseteq [n]$,
    $\sfor_{P}$ is the disjunction of the $i$-th arguments, $i\in P$).
    One verifies easily that $T_{\path}$ is a preclone, which is
    generated by the binary $\sfor_{2}$ function, the nullary constants
    $\true_{0}$ and $\false_{0}$, and the unary constant $\false_{1}$.
    The rank 1 elements of $T_{\path}$ form a 3-element monoid,
    isomorphic to the multiplicative monoid $\{1,a,b\}$ with $xy = y$
    for $x,y\ne 1$, known as $U_{2}$ in the literature on monoid
    theory, e.g. \cite{Pin}.
\end{expl}

%%%%%%%%%%%%%%%%%%%%%%
\subsection{Representation of preclones}\label{sec representation}

Section \ref{preclone vs tree automata} shows the importance of the
representation of preclones as preclones of transformations. It is not
difficult to establish the following analogue of Cayley's theorem.

\begin{prop}\label{prop-rep1}
    Every preclone can be embedded in a preclone of transformations.
\end{prop}

\proof
Suppose that $S$ is a preclone and let $Q$ be the disjoint union of
the sets $S_n$, $n \geq 0$. For each $f \in S_n$, let $\ol{f}$ be the
function $Q^n \to Q$ given by
\begin{eqnarray*}
\ol{f}(g_1,\ldots,g_n) &=& f \cdot (g_1 \oplus \cdots \oplus g_n).
\end{eqnarray*} 
The assignment $f \mapsto \ol{f}$ defines an injective morphism 
$S \to T(Q)$.
\eop

This result however is not very satisfactory: it does not tell us
whether a finitary preclone can be embedded in the preclone of
transformations of a finite set. It is actually not always the case,
and this leads to the following discussion.

Let $k \geq 0$. We say that a preclone $S$ is \emph{$k$-determined} if
distinct elements can be separated by $k$-ary equations. Formally, let
$\sim_k$ denote the following equivalence relation: for all $f,g \in
S_n$ ($n\ge 0$),
\begin{eqnarray*}
    f\sim_k g & \Longleftrightarrow & f\cdot h = g\cdot h, \hbox{ for
    all $h\in S_{n,\ell}$ with $\ell \leq k$.}
\end{eqnarray*}
Note that for each $\ell \leq k$, $\sim_k$ is the identity relation on
$T_{\ell}$. We call $S$ \emph{$k$-determined} if the relation $\sim_k$
is the identity relation on each $S_n$, $n \geq 0$, and we say that $S$
is \emph{finitely determined} if it is $k$-determined for some integer
$k$. We also say that a $pg$-pair $(S,A)$ is $k$-determined (resp. finitely
determined) if $S$ is.

\begin{expl}
    The preclone of transformations of a set is 0-determined.
\end{expl}

\note{Several important questions:
(1)
is it true that finitary does not imply embeddable in $T(Q)$, $Q$
finite?
(2)
Is $T(Q)$, $Q$ finite, finitely generated?
(2$'$)
Is there a cardinality bound (say, exponential) for the rank n
elements of a finitely generated preclone?
(3)
Are there finitary, not finitely determined preclones?
(4)
Is it true that an effectively given preclone (a fg subpreclone
of $T(Q)$ for some finite $Q$) is necessarily finitely
determined?}

We observe the two following easy lemmas.

\begin{lem} 
For each $k$, $\sim_k$ is a congruence relation. 
\end{lem}

\proof
Let $f,g\in S_{n}$ be $\sim_{k}$-equivalent. For each $i \in [n]$, let
$f_i,g_{i}\in S_{m_{i}}$, such that $f_{i} \sim_k g_i$. We want to show
that $f\cdot(f_1 \oplus \cdots \oplus f_n) \sim_{k}g \cdot (g_1 \oplus
\cdots \oplus g_n)$.

Let $m = \sum_{i\in[n]}m_{i}$ and let $h\in S_{m,\ell}$ for some
$\ell\leq k$. Then $h$ is an $m$-tuple, and we let $h_{1}$ be the tuple
of the first $m_{1}$ terms of $h$, $h_{2}$ consist of the next $m_{2}$
elements, etc, until $h_{n}$, which consists of the last $m_{n}$
elements of $h$. Note that each $h_{i}$ lies in some
$S_{m_{i},\ell_{i}}$ and that $\sum_{i\in[n]}\ell_{i} = \ell$. In
particular, $\ell_{i}\leq k$ for each $i$ and we have
\begin{eqnarray*}
f \cdot (f_1 \oplus \cdots \oplus f_n) \cdot h 
&=& 
f \cdot (f_1 \cdot h_1 \oplus \cdots \oplus f_n\cdot h_n)\\
&=& 
f \cdot (g_1 \cdot h_1 \oplus \cdots \oplus g_n\cdot h_n)\\
&=& 
g \cdot (g_1 \cdot h_1 \oplus \cdots \oplus g_n\cdot h_n)\\
&=&
g \cdot (g_1 \oplus \cdots \oplus g_n) \cdot h.  
\end{eqnarray*}  
\eop

\begin{lem}\label{simk k determined}
    For each $k \geq 0$, the quotient preclone $S/{\sim_{k}}$
    is $k$-determined.
\end{lem}

\proof
Let $T = S/{\sim_{k}}$ and let $[f],[g]\in T_{n}$, where $[f]$ denotes
the $\sim_k$-equivalence class of $f$ (necessarily in $S_{n}$). Let
$\ell\leq k$ and assume that $[f]\cdot [h] = [g]\cdot [h]$ for each
$h\in T_{n,\ell}$. Then $ f \cdot h \sim_k g \cdot h$ for each $h$. But
$f\cdot h$ and $g\cdot h$ lie in $S_{\ell}$, and we already noted that
$\sim_{k}$ is the identity relation on $S_{\ell}$ (since $\ell\leq k$).
Thus $f \cdot h = g \cdot h$ for all $h\in S_{n,\ell}$, and since this
holds for each $\ell \leq k$, we have $f \sim_k g$, and hence $[f] =
[g]$.
\eop 

We say that a preclone morphism $\phi\colon S \to T$ is
\emph{$k$-injective} if it is injective on each $S_{\ell}$ with $\ell
\leq k$. The next lemma, relating $k$-determination and $k$-injectivity,
will be used to discuss embeddability of a finitary preclone in the
preclone of transformations of a finite set.

\begin{lem}\label{prop-k-injective}
    Let $S$ be a $k$-determined preclone. If $\phi\colon S \to T$ is a
    $k$-injective morphism, then $\phi$ is injective.
\end{lem} 

\proof 
If $\phi(f) = \phi(g)$ for some $f,g \in S_n$, then $\phi(f\cdot h) =
\phi(f) \cdot \phi(h) = h(g) \cdot \phi(h) = \phi(g \cdot h)$, for all
$h\in S_{n,\ell}$ with $\ell \leq k$. Since $\phi$ is $k$-injective, it
follows $f \cdot h = g \cdot h$ for all $h\in S_{n,\ell}$ with
$\ell\leq k$, and since $S$ is $k$-determined, this implies $f = g$.
\eop

\begin{prop}\label{prop-emb-finite}
    Let $S$ be a finitary and finitely determined preclone. Then there
    is a finite set $Q$ such that $S$ embeds in $T(Q)$. If in
    addition $S$ is 0-determined, the set $Q$ can be taken equal to
    $S_{0}$.
\end{prop} 

\proof 
We modify the construction in the proof of Proposition~\ref{prop-rep1}.
Let $k\ge 0$ be such that $S$ is $k$-determined, and let $Q = \{\bot\}
\cup \bigcup_{i \leq k}S_i$, where the sets $S_i$ are assumed to be
pairwise disjoint and $\bot$ is a new symbol, not in any of those sets.
For each $f\in S_n$ ($n\ge 0$), let $\phi(f) = \ol{f}: Q^n \to Q$ be
the function defined by
$$\ol{f}(q_1,\ldots,q_n)
= \cases{
f \cdot (q_1 \oplus \cdots \oplus q_n)
&  if $q_1 \in S_{m_1},\ldots ,q_n\in S_{m_n}$ and $\sum m_i \leq
k$,\cr
\bot & otherwise.}$$ 
It is easy to check that $\phi$ is a morphism. By
Lemma~\ref{prop-k-injective}, $\phi$ is injective.

To conclude, we observe that if $k = 0$, we can choose $Q = S_{0}$
since $\sum_{i=1}^n m_{i}\leq 0$ is possible if and only if each $m_{i}
= 0$.
\eop  

For later use we also note the following technical results.

\begin{prop}\label{prop-extension}
    Let $S$ and $T$ be preclones, with $T$ $k$-determined. Let $G$ be a
    (ranked) generating set of $S$ and let $\phi\colon G\rightarrow
    T$ be a rank-preserving map, whose range includes all of $T_\ell$,
    for each $\ell \leq k$. Then $\phi$ can be extended to a
    preclone morphism $\ol{\phi}\colon S \to T$ iff for all $g \in
    G_n$, $n \geq 0$, and for all $h\in G_{n,\ell}$ with $\ell \leq k$,
    \begin{eqnarray}
	\phi(g \cdot h) &=& \phi(g)\cdot
	\phi(h).\label{eq-extension}
    \end{eqnarray}
\end{prop}

\proof
Condition~(\ref{eq-extension}) is obviously necessary, and we show that
it is sufficient.

Let $f\in S_{n}$, $n\ge 0$. Any possible image of $f$ by a preclone
morphism is an element $g\in T_{n}$ such that, if $h \in T_{n,\ell}$
for some $\ell \leq k$ and if $h' \in G_{n,\ell}$ is such that
$\phi(h') = h$, then $g\cdot h = \phi(f\cdot h')$. Since $T$ is
$k$-determined and each $T_{\ell}$ ($\ell\leq k$) is in the range of
$\phi$, the element $g$ is completely determined by $f$. That is, if
an extension of $\phi$ exists, then it is unique.

We now show the existence of this extension. We want to assign an image
to an arbitrary element $f$ of $S$, and we proceed by induction on the
height of an expression of $f$ in terms of the elements of $G$; such an
expression exists since $G$ generates $S$. If $f\in G$, we let
$\ol\phi(f) = \phi(f)$. Note also that if $f = \1$, then we let
$\ol\phi(f) = \1$. If $f\not\in G$, then $f = g\cdot h$ for some $g\in
G\cap S_{n}$ and some $h = h_{1}\oplus \cdots \oplus h_{n}$. By
induction, the elements $\ol\phi(h_i)$ are well defined for each $i \in
[n]$. We then let $\ol\phi(f) =
\phi(g)\cdot(\ol\phi(h_{1})\oplus\cdots\oplus\ol\phi(h_{n}))$.

To show that $\ol\phi(f)$ is well-defined, we consider a different
decomposition of $f$, say, $f = g'\cdot h'$ with $h' =
h'_{1}\oplus\cdots\oplus h'_{m}$.
If $f$ has rank $\ell\leq k$, then $h\in S_{n,\ell}$ and $h'\in
S_{m,\ell}$, so $h = \phi(\bar h)$ and $h' = \phi(\bar h')$ for some
$\bar h\in G_{n,\ell}$ and $\bar h'\in G_{m,\ell}$. By
Condition~(\ref{eq-extension}), we have
$$\phi(g)\cdot \ol\phi(\bar h) = \phi(g)\cdot\phi(\bar h) =
\phi(g\cdot h) = \phi(f),$$
and by symmetry, $\phi(g)\cdot\ol\phi(\bar h) =
\phi(g')\cdot\ol\phi(\bar h')$. So $\ol\phi$ is well defined on all 
the elements of $S$ of rank at most $k$.

Now if $f$ has rank $\ell > k$, let $x\in T_{\ell,p}$ with $p\leq k$.
Then there exists $x'\in G_{\ell,p}$ such that $x = \phi(x')$. Note
that $h\cdot x'$ and $h'\cdot x'$ are well defined, in $S_{n,p}$ and in
$S_{m,p}$ respectively. In particular, $\ol\phi(h\cdot x')$ is
well defined, and equal to $\ol\phi(h)\cdot x$. Similarly,
$\ol\phi(h'\cdot x')$ is well defined, equal to $\ol\phi(h')\cdot
x$. It follows that
$$(\phi(g)\cdot\ol\phi(h))\cdot x = \phi(g)\cdot(\ol\phi(h)\cdot x)
= \phi(g)\cdot\ol\phi(h\cdot x') = \ol\phi(g\cdot(h\cdot x')) =
\ol\phi(f\cdot x').$$
By symmetry, $(\phi(g')\cdot\ol\phi(h'))\cdot x =
(\phi(g)\cdot\ol\phi(h))\cdot x$, and since $T$ is $k$-determined, we
have $\phi(g)\cdot\ol\phi(h) = \phi(g')\cdot\ol\phi(h')$. Thus,
$\ol\phi$ is well defined on $S$.

By essentially the same argument, one verifies that $\ol{\phi}$
preserves composition.
\eop 

\begin{cor}\label{cor-extension}
    Let $S$ and $T$ be $k$-determined preclones that are generated by
    their elements of rank at most $k$. If there exist bijections from
    $S_\ell$ to $T_\ell$ for each $\ell \leq k$, that preserve all
    compositions of the form $f \cdot g$, where $f \in S_n$, $g \in
    S_{n,\ell}$ with $n,\ell \leq k$, then $S$ and $T$ are isomorphic.
\end{cor} 

\proof
Using the fact that $S$ is generated by its elements of rank at most
$k$ and $T$ is $k$-determined, Proposition~\ref{prop-extension} shows
that the given bijections extend to a morphism from $S$ to $T$. This
morphism is onto since $T$ as well is generated by its rank $k$
elements. It is also $k$-injective by construction, and hence it is
injective by Lemma~\ref{prop-k-injective} since $S$ is $k$-determined.
\eop

%%%%%%%%%%%%%%%%%%%%%%
\subsection{Preclones, magmoids and strict monoidal categories}
\label{sec magmoids}

The point of this short subsection is to verify the close connection
between the category of preclones and the category of \emph{magmoids},
cf. \cite{Arnoldetal}, which are in turn a special case of \emph{strict
monoidal categories}, cf. \cite{MacLane}. We recall that a magmoid is a
category $M$ whose objects are the nonnegative integers equipped with
an associative bifunctor $\oplus$ such that $0\oplus x = x = x\oplus
0$. A morphism of magmoids is a functor that preserves objects and
$\oplus$. We say that a magmoid $M$ is \emph{determined by its scalar
morphisms} if each morphism $f\colon n\to m$ can be written in a unique
way as a $\oplus$-sum $f_1 \oplus \cdots \oplus f_n$, where each $f_i$
is a morphism with source $1$. Moreover, there is a morphism $0$ to $n$
if and only if $n = 0$ (in which case there is a unique morphism).

\begin{prop}
    The category of preclones is equivalent to the full subcategory 
    of magmoids spanned by those magmoids which are determined by their 
    scalar morphisms. 
\end{prop} 

\proof
With each preclone $S$, we associate a category whose objects are the
nonnegative integers and whose morphisms $n \to m$ are the elements of
$S_{n,m}$, that is, the $n$-tuples of elements of $S$ of total rank
$m$. Composition is defined in the following way: let $f = f_1\oplus
\cdots \oplus f_n \in S_{n,m}$ and $g = g_1\oplus \cdots \oplus g_m\in
S_{m,p}$, and suppose that $f_i\in S_{m_i}$, $i \in [n]$ (so that $m =
\sum_{i\in[n]}m_{i}$). For each $i$, let $\ol{g}_i = g_{m_1+ \cdots +
m_{i-1}+1} \oplus \cdots \oplus g_{m_1+\cdots + m_{i}}$. Then we let
\begin{eqnarray*}
f \cdot g &=& f_1 \cdot \ol{g}_1 \oplus \cdots \oplus f_m\cdot \ol{g}_m. 
\end{eqnarray*} 
The identity morphism at object $n$ is the $n$-tuple ${\bf n} = \1
\oplus \cdots \oplus \1$. Note that when $n = 0$, this is the unique
morphism $0 \to 0$, and there are no morphisms from $0$ to $n$ if $n
\ne 0$.

One may then regard $\oplus$ as a bifunctor $S \times S \to S$ that
maps a pair $(f,g) $ with $f = f_1 \oplus\cdots \oplus f_n \in S_{n,p}$
and $g = g_1\oplus \cdots \oplus g_m \in S_{m,q}$ to the morphism $f_1
\oplus \cdots \oplus f_n \oplus g_1\oplus \cdots \oplus g_m$ from $n+m$
to $p+q$. Then $S$, equipped with the bifunctor $\oplus$, is 
a magmoid. Moreover, $S$, as a magmoid, is determined by its scalar morphisms. 

It is clear that each preclone morphism determines a functor between
the corresponding magmoids which is the identity function on objects
and preserves $\oplus$ and is thus a morphism of magmoids. 

Conversely, if $M$ is a magmoid determined by its scalar morphism,
then its morphisms with sourse $1$ constitute a preclone $S$,
moreover, $M$ is isomorphic to the magmoid determined by $S$. 
\eop

%%%%%%%%%%%%%%%%%%%%%%%%%%%%%%%%%%%%%%
\section{Recognizable tree languages}\label{grand sec recog}

As discussed in the introduction, the theory of (regular) tree
languages is well developped \cite{GecsegSteinby,GSHB,tata} (see also
Section~\ref{preclone vs tree automata} above). Here we slightly extend
the notion of tree languages, to mean any subset of some $\Sigma
M_{k}$, $k\geq 0$. In the classical setting, tree languages are subsets
of $\Sigma M_{0}$.

The preclone structure on $\Sigma M$, described in
Section~\ref{algebraic}, leads in a standard fashion to a definition of
recognizable tree languages
\cite{EilenbergWright,MezeiWright,Courcelle,CourcelleWeil,WeilMFCS}.
This is discussed in some detail in Section~\ref{sec recog}. As we will
see in Section~\ref{sec regular tree}, recognizability extends the
classical notion of regularity for tree languages, and it gives us
richer algebraic tools to discuss these languages. Further examples are
given in Section~\ref{sec more examples trees}.

%%%%%%%%%%%%%%%%%%%%%%%%%%%%%%%%%%%%%%
\subsection{Syntactic preclones}\label{sec recog}

Suppose that $\alpha: \Sigma M \to S$ is a preclone morphism, or a
morphism $(\Sigma M,\Sigma) \to (S,A)$. We say that a subset $L$ of
$\Sigma M_{k}$ is \emph{recognized by $\alpha$} if $L =
\alpha\inv\alpha(L)$, or equivalently, if $L = \alpha\inv(F)$ for some
$F \subseteq S_k$. Moreover, we say that $L$ is recognized by $S$, or
by $(S,A)$, if $L$ is recognized by some morphism $\Sigma M \to S$ or
$(\Sigma M,\Sigma) \to (S,A)$. Finally, we say that a subset $L$ of
$\Sigma M_{k}$ is \emph{recognizable} if it is recognized by a finitary
preclone, or $pg$-pair. As usual, the notion of recognizability can be
expressed equivalently by stating that $L$ is \emph{saturated} by some
\emph{locally finite} congruence on $\Sigma M$, that is, $L$ is a union
of classes of a congruence which has finite index on each sort
\cite{Courcelle1996,CourcelleHdbook,WeilMFCS}.

With every subset $L\subseteq\Sigma M_{k}$, recognizable or not, we
associate a congruence on $\Sigma M$, called the
\textit{syntactic congruence} of $L$. This relation is defined
as follows. First, an \textit{$n$-ary context in $\Sigma M_{k}$} is a
tuple $(u,k_1,v,k_2)$ where
\begin{itemize}
   \item $k_{1}, k_{2}$ are nonnegative integers,
   
   \item $u  \in \Sigma M_{k_1 + 1 + k_2}$, and
   
   \item $v = v_1 \oplus \cdots \oplus v_n \in \Sigma
M_{n,\ell}$, with $k = k_1 + \ell + k_2$.
\end{itemize}
$(u,k_1,v,k_2)$ is an \textit{$L$-context} of an element $f\in \Sigma
M_n$ if $u\cdot (\k_1 \oplus f \cdot v \oplus \k_2) \in L$. Recall that
$\k$ denotes the $\oplus$-sum of $k$ terms equal to $\1$. Below, when
$k_1$ and $k_2$ are clear from the context (or do not play any role),
we will write just $(u,v)$ to denote the context $(u,k_1,v,k_2)$.

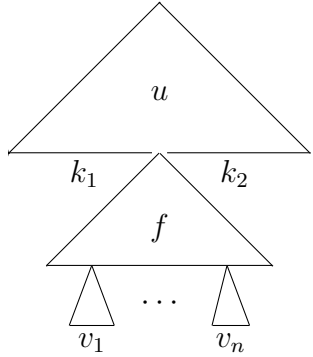
\begin{center}
\begin{picture}(49,44)(0,-44)
% \put(0,-45){\framebox(49,45){}}
    \node[Nw=0.1,Nh=0.0,Nmr=0.0](n0)(24.0,0.0){}

    \node[Nw=0.0,Nh=0.0,Nmr=0.0](n1)(4.0,-20.0){}

    \node[Nw=0.0,Nh=0.0,Nmr=0.0](n2)(44.0,-20.0){}

    \node[Nframe=n,Nw=2.0,Nh=2.0,Nmr=0.0](n3)(24.0,-20.0){}

    \node[Nframe=n,NLangle=0.0](n4)(24.0,-12.0){$u$}

    \drawedge[AHnb=0](n0,n1){ }

    \drawedge[AHnb=0](n0,n2){ }

    \drawedge[AHnb=0](n3,n1){$k_{1}$}

    \drawedge[AHnb=0](n2,n3){$k_{2}$}

    \node[Nframe=n,Nw=0.0,Nh=0.0,Nmr=0.0](n5)(24.0,-20.0){}

    \node[Nw=0.0,Nh=0.0,Nmr=0.0](n6)(9.0,-35.0){}

    \node[Nw=0.0,Nh=0.0,Nmr=0.0](n7)(39.0,-35.0){}

    \node[Nframe=n,NLangle=0.0](n11)(24.0,-30.0){$f$}

    \drawedge[AHnb=0](n5,n6){ }

    \drawedge[AHnb=0](n6,n7){ }

    \drawedge[AHnb=0](n7,n5){ }

%     \node[Nframe=n,NLangle=0.0](n4)(24.0,-12.0){$u$}

    \node[Nframe=n,Nw=0.0,Nh=0.0,Nmr=0.0](n8)(15.0,-35.0){}

    \node[Nframe=n,Nw=0.0,Nh=0.0,Nmr=0.0](n9)(12.0,-43.0){}

    \node[Nframe=n,Nw=0.0,Nh=0.0,Nmr=0.0](n10)(18.0,-43.0){}

    \drawedge[AHnb=0](n8,n10){ }

    \drawedge[AHnb=0](n10,n9){$v_{1}$}

    \drawedge[AHnb=0](n9,n8){}
    
    \node[Nframe=n,Nw=0.0,Nh=0.0,Nmr=0.0](n15)(24.0,-40.0){$\cdots$}

    \node[Nframe=n,Nw=0.0,Nh=0.0,Nmr=0.0](n12)(33.0,-35.0){}

    \node[Nframe=n,Nw=0.0,Nh=0.0,Nmr=0.0](n13)(31.0,-43.0){}

    \node[Nframe=n,Nw=0.0,Nh=0.0,Nmr=0.0](n14)(36.0,-43.0){}

    \drawedge[AHnb=0](n12,n14){ }

    \drawedge[AHnb=0](n14,n13){$v_{n}$}

    \drawedge[AHnb=0](n13,n12){}

\end{picture}
\end{center}

For each $f,g\in \Sigma M_n$, we let $f\sim_L g$ if and only if
$f$ and $g$ have the same $L$-contexts.

\begin{prop}
      The relation $\sim_{L}$, associated with a subset $L$ of $\Sigma
      M_k$, is a preclone congruence which saturates $L$.
\end{prop}

\proof 
Suppose that $f,f' \in \Sigma M_n$ and $g,g' \in \Sigma M_{n,m}$ with
$f \sim_L f'$, $g = g_{1}\oplus\cdots\oplus g_{n}$, $g' =
g'_{1}\oplus\cdots\oplus g'_{n}$ and $g_i \sim_L g'_i$ for each $1 \leq
i \leq n$. We prove that $f\cdot g \sim_L f'\cdot g'$. Let $m_{i}$ be
the rank of $g_{i}$ and $g'_{i}$, so that $m = \sum_{i\in[n]}m_i$ and
consider any $m$-ary context $(u,k_{1},v,k_{2})$ in $\Sigma M_{k}$.
Then $v\in \Sigma M_{m,\ell}$ with $\ell = k-(k_{1}+k_{2})$. Thus, $v$
is an $m$-tuple, and we let $w_{1}$ be the $\oplus$-sum of the first
$m_{1}$-terms of $v$, $w_{2}$ be the $\oplus$-sum of the following
$m_{2}$-terms of $v$, etc, until finally $w_{n}$ is the $\oplus$-sum of
the last $m_{n}$ terms of $v$. In particular, we may write $v =
w_{1}\oplus\cdots\oplus w_{n}$.

Since $f\sim_{L} f'$, we have
$$u \cdot (\k_1 \oplus f \cdot g\cdot v
\oplus \k_2) \in L \Longleftrightarrow u \cdot (\k_1 \oplus f' \cdot
g\cdot v \oplus \k_2) \in L.$$
It suffices to consider the $n$-ary context $(u,\k_{1},g\cdot
v,\k_{2})$, where $g \cdot v = (g_1 \oplus \cdots \oplus g_n)\cdot v$
stands for $g_{1}\cdot w_{1}\oplus \cdots \oplus g_{n}\cdot w_{n}$.

Moreover, since $g_i \sim_L g'_i$, we have
\begin{eqnarray*}
    \lefteqn{ u \cdot (\k_1 \oplus f\cdot (g'_1 \cdot w_1 \oplus \cdots
    \oplus g'_{i-1}\cdot w_{i-1} \oplus g_i \cdot w_i \oplus g_{i+1}
    \cdot w_{i+1}\cdots \oplus g_n \cdot w_n) \oplus \k_2) \in L }\\
    &\Longleftrightarrow & 
    u \cdot (\k_1 \oplus f\cdot (g'_1 \cdot w_1 \oplus \cdots \oplus
    g'_i \cdot w_i \oplus g_{i+1} \cdot w_{i+1} \oplus \cdots \oplus
    g_n \cdot w_n) \oplus \k_2) \in L,
\end{eqnarray*}
for each $1 \leq i \leq n$. To justify this statement, it suffices to
consider the following $m_{i}$-ary context (for $g_{i}$ and $g'_{i}$),
$$\Big(u\cdot(g'_{1}\cdot w_{1} \oplus \cdots g'_{i-1}\cdot w_{i-1}
\oplus \1 \oplus g_{i+1}\cdot w_{i+1} \oplus \cdots \oplus g_{n}\cdot
w_{n}),\ \k_{1}+\ellbf_{i,1},\ w_{i},\ \ellbf_{i,2}+\k_{2} \Big),$$
where $\ell_{i,1}$ is the sum of the ranks of $w_{1},\ldots,w_{i-1}$,
and $\ell_{i,2}$ is the sum of the ranks of $w_{i+1},\ldots,w_{n}$.

% \insertpicture

We now have 
\begin{eqnarray*}
    u \cdot (\k_1 \oplus f' \cdot g\cdot v \oplus \k_2) \in L
    &\Longleftrightarrow & u \cdot (\k_1 \oplus f \cdot (g_1 \cdot w_1
    \oplus \cdots \oplus g_n\cdot w_n) \oplus \k_2) \in L\\
    &\Longleftrightarrow & u \cdot (\k_1 \oplus f \cdot (g'_1 \cdot w_1
    \oplus \cdots \oplus g_n\cdot w_n) \oplus \k_2) \in L\\
    &\vdots & \\
    &\Longleftrightarrow & u \cdot (\k_1 \oplus f \cdot (g'_1 \cdot w_1
    \oplus \cdots \oplus g'_n\cdot w_n) \oplus \k_2) \in L\\
    &\Longleftrightarrow & u \cdot (\k_1 \oplus f \cdot g'\cdot v
    \oplus \k_2) \in L.
\end{eqnarray*}
This completes the proof that $\sim_L$ is a congruence. Next we observe
that an element $f\in\Sigma M_{k}$ is in $L$ if and only if the $k$-ary
context $(\1,\k)$ is an $L$-context of $f$: it follows immediately that
$\sim_L$ saturates $L$.
\eop 

We denote by $(M_L,\Sigma_L)$ the quotient $pg$-pair $(\Sigma
M/{\sim_L}, \Sigma/{\sim_L})$, called the \textit{syntactic $pg$-pair}
of $L$. $M_L$ is the \textit{syntactic preclone} of $L$ and the
projection morphism $\eta_{L}\colon\Sigma M\rightarrow M_{L}$, or
$\eta_{L}\colon(\Sigma M, \Sigma) \rightarrow (M_{L}, \Sigma_L)$, is
the \textit{syntactic morphism} of $L$. We note the following, expected
result.

\begin{prop}\label{cor-synt morph} 
      The syntactic congruence of a subset $L$ of $\Sigma M_k$ is the
      coarsest preclone congruence which saturates $L$. A preclone
      morphism $\alpha\colon \Sigma M\rightarrow S$ (resp. a morphism
      of \pgs\ $\alpha\colon (\Sigma M,\Sigma)\rightarrow (S,A)$)
      recognizes $L$ if and only if $\alpha$ can be factored through
      the syntactic morphism $\eta_{L}$. In particular, $L$ is
      recognizable if and only if $\sim_{L}$ is locally finite, if and
      only if $M_{L}$ is finitary.
\end{prop}

\proof 
Let $\approx$ be a congruence saturating $L$ and assume that $f,g\in
\Sigma M_{n}$ are $\approx$-equivalent. Let $(u,k_{1},v,k_{2})$ be an
$n$-ary context: then $u\cdot (\k_1 \oplus f \cdot v \oplus \k_2)
\approx u\cdot (\k_1 \oplus g \cdot v \oplus \k_2)$, and since
$\approx$ saturates $L$, $u\cdot (\k_1 \oplus f \cdot v \oplus \k_2)
\in L$ iff $u\cdot (\k_1 \oplus g \cdot v \oplus \k_2) \in L$. Since
this holds for all $n$-ary contexts in $\Sigma M_{k}$, it follows that
$f \sim_L g$.
\eop

We also note that syntactic preclones are finitely determined.

\begin{prop}\label{syntactic finitely determined}
    The syntactic preclone of a subset $L$ of $\Sigma M_k$ is
    $k$-determined.
\end{prop}

\proof 
We show that if $f,g \in \Sigma M_{n}$ and $f \cdot h \sim_L g \cdot h$
for all $h\in \Sigma M_{n,\ell}$ with $\ell \leq k$, then $f \sim_L g$,
that is, $f$ and $g$ have the same $L$-contexts.

Let $(u,k_{1},v,k_{2})$ be an $L$-context of $f$. Note in particular
that $v\in \Sigma M_{n,p}$ with $k = k_{1} + p + k_{2}$. It follows
that $f\cdot v\in \Sigma M_{p}$, and that $f\cdot v \sim_{L} g\cdot v$.
Moreover $(u,k_{1},\p,k_{2})$ is an $L$-context of $f\cdot v$. But in
that case, $(u,k_{1},\p,k_{2})$ is also an $L$-context of $g\cdot v$,
and hence $(u,k_{1},v,k_{2})$ is an $L$-context of $g$, which concludes
the proof.
\eop

%%%%%%%%%%%%%%%%%%%%%%%%%%%%
\subsection{The usual notion of regular tree languages}\label{sec
regular tree}

We now turn to tree languages in the usual sense, that is, subsets of
$\Sigma M_{0}$. For these sets, there exists a well-known notion of
(bottom-up) tree automaton, whose expressive power is equivalent to
monadic second-order definability, to certain rational expressions, and
to recognizability by a finite $\Sigma$-algebra
\cite{GecsegSteinby,GSHB} (see
Section~\ref{preclone vs tree automata}). The tree languages captured
by these mechanisms are said to be \emph{regular}. It is an essential
remark (Theorem~\ref{syntactic gp} below) that the regular tree
languages are exactly the subsets of $\Sigma M_{0}$ that are recognized
by a finitary preclone.

Recall that the minimal tree automaton of a regular tree language is
the least deterministic tree automaton accepting it, and the
$\Sigma$-algebra associated with this automaton is called the
\emph{syntactic $\Sigma$-algebra} of the language. It is characterized
by the fact that the natural morphism from the initial $\Sigma$-algebra
to the syntactic $\Sigma$-algebra of $L$, factors through every
morphism of $\Sigma$-algebra which recognizes $L$ (see
Section~\ref{preclone vs tree automata} and
\cite{GecsegSteinby,GSHB,Almeida}).

\begin{thm}\label{syntactic gp}
    A tree language $L \subseteq \Sigma M_0$ is recognizable if and
    only if it is regular. Moreover, the syntactic preclone (resp.
    $pg$-pair) of $L$ is the preclone (resp. $pg$-pair) associated with
    its syntactic $\Sigma$-algebra.
\end{thm}

\proof
Let $Q$ be the syntactic $\Sigma$-algebra of $L$, let $(S,A)$ be its
syntactic $pg$-pair, and let $\eta\colon (\Sigma M,\Sigma)\rightarrow
(S,A)$ be its syntactic morphism. As discussed in Section~\ref{preclone
vs tree automata}, the $pg$-pair associated with $Q$, $\pg(Q)$,
recognizes $L$, and hence the syntactic morphism of $L$ factors through
an onto morphism of \pgs\ $\pg(Q) \rightarrow (S,A)$. In particular, if
$L$ is regular, then $Q$ is finite, so $\pg(Q)$ is finitary, and so is
$(S,A)$: thus $L$ is recognizable.

Conversely, assume that $L$ is recognizable. Since $(S,A)$ is finitary
and 0-determined (Proposition~\ref{syntactic finitely determined}), so
$(S,A)$ is isomorphic to a sub-$pg$-pair of $T(S_{0})$ by
Proposition~\ref{prop-emb-finite}. Using again the discussion in
Section~\ref{preclone vs tree automata}, $S_{0}$ has a natural
structure of $\Sigma$-algebra (via the morphism $\eta$), such that
$(S,A) = \pg(S_{0})$ and such that $S_{0}$ recognizes $L$ as a
$\Sigma$-algebra. In particular, $L$ is recognized by a finite
$\Sigma$-algebra, and hence $L$ is regular.

Moreover, the recognizing morphism $\Sigma M_{0}\rightarrow S_{0}$ is
the restriction to $\Sigma M_{0}$ of $\eta$, the syntactic morphism of
$L$. Therefore there exists an onto morphism of $\Sigma$-algebras
$S_{0}\rightarrow Q$, which in turn induces a morphism of preclones
from $(S,A) = \pg(S_{0})$ onto $\pg(Q)$. Since $Q$ and $S_{0}$ are
finite, it follows that the morphisms between them described above are
isomorphisms, and this implies that $\pg(Q)$ is isomorphic to $(S,A)$.
\eop 

While not difficult, Theorem~\ref{syntactic gp} is important because it
shows that we are not introducing a new class of recognizable tree
languages. We are simply associating with each regular tree language a
finitary algebraic structure which is richer than its syntactic
$\Sigma$-algebra (a.k.a. minimal deterministic tree automaton). This
theorem also implies that the syntactic $pg$-pair of a recognizable
tree language has an effectively computable finite presentation.

\begin{remark}
    If $L\subseteq \Sigma M_{0}$, the definition of the syntactic
    congruence of $L$ involves the consideration of $n$-ary contexts in
    $\Sigma M_{0}$. Such contexts are necessarily of the form
    $(u,\0,v,\0)$, where $u\in\Sigma M_{1}$ and $v\in\Sigma M_{n,0}$,
    which somewhat simplifies matters.
\end{remark}    
 
%%%%%%%%%%%%%%%%%%%%%%%%%%%%
\subsection{More examples of recognizable tree languages}\label{sec
more examples trees}

The examples in this section are directly related with the preclones
discussed in Section~\ref{simple examples}. Let $\Delta$ be a ranked
\emph{Boolean} alphabet, that is, a ranked alphabet such that each
$\Delta_{n}$ is either empty or equal to $\{0_{n},1_{n}\}$, and
$\Delta_{0}$ and at least one $\Delta_{n}$ ($n\geq 2$) are nonempty.
Let $k\geq 0$ be an integer.

\paragraph*{Verifying the occurrence of a letter}
Let $K_{k}(\exists)$ be the set of all trees in $\Delta M_k$ containing
at least one vertex labeled $1_{n}$ (for some $n$). Then
$K_{k}(\exists)$ is recognizable, by a morphism into the preclone
$T_{\exists}$ (see Example~\ref{define Texists}).

Let $\alpha\colon \Delta M\to T_{\exists}$ be the morphism of preclones
given by $\alpha(0_{n}) = \sfor_{n}$ ($\alpha(0_{0}) = \false_{0}$) and
$\alpha(1_{n}) = \true_{n}$ whenever $\Delta_{n}\ne\emptyset$. It is
not difficult to verify that $\alpha\inv(\true_{k}) =
K_{k}({\exists})$. Moreover, $\alpha(\Delta)$ contains a generating set
of $T_{\exists}$, so $\alpha$ is onto, and the syntactic morphism of
$K_{k}(\exists)$ factors through $\alpha$. But $T_{\exists}$ has at
most 2 elements of each rank, so any proper quotient $M$ of
$T_{\exists}$ has exactly one element of rank $n$ for some integer $n$.
One can then show that $M$ cannot recognize $K_{k}(\exists)$. Thus the
syntactic $pg$-pair of $K_{k}(\exists)$ is
$(T_{\exists},\alpha(\Delta))$.

If $\Sigma$ is any ranked alphabet such that $\Sigma_{0}$ and at least
one $\Sigma_{n}$ ($n>1$) is nonempty, if $\Sigma'$ is a proper nonempty
subset of $\Sigma$, and $K_{k}(\Sigma')$ is the set of all trees in
$\Sigma' M_{k}$ containing at least a node labeled in $\Sigma'$, then
$K_{k}(\Sigma')$ too has syntactic preclone $T_{\exists}$. The
verification of this fact can be done using a morphism from $\Sigma M$
to $\Delta M$, mapping each letter $\sigma$ of rank $n$ to $1_{n}$ if
it is in $\Sigma'$, to $0_{n}$ otherwise.

\paragraph*{Counting the occurrences of a letter}
Let $p,r$ be integers such that $0\leq r < p$ and let
$K_{k}(\exists^r_p)$ consist of the trees in $\Delta M_k$ such that
the number of vertices labeled $1_{n}$ (for some $n$) is congruent to
$r$ modulo $p$. Then $K_{k}(\exists^r_p)$ is recognizable, by a
morphism into the preclone $T_{p}$ (see Example~\ref{define Tp}).

Let indeed $\alpha\colon \Delta M\to T_{p}$ be the morphism given by
$\alpha(0_{n}) = f_{n,0}$ and $\alpha(1_{n}) = f_{n,1}$ whenever
$\Delta_{n}\ne\emptyset$. Then one verifies that $\alpha\inv(f_{k,r}) =
K_{k}({\exists^r_p})$. Moreover, $\alpha(\Delta)$ contains a generating
set of $T_{p}$, so $\alpha$ is onto, and the syntactic morphism of
$K_{k}(\exists^r_p)$ factors through $\alpha$. An elementary
verification then establishes that no proper quotient of $T_{p}$ can
recognize $K_{k}(\exists^r_p)$, and hence the syntactic $pg$-pair of
$K_{k}(\exists^r_p)$ is $(T_{p},\alpha(\Delta))$.

As above, this can be extended to recognizing the set of all trees in $\Sigma
M_{k}$ where the number of nodes labeled in some proper nonempty
subset $\Sigma'$ of $\Sigma$ is congruent to $r$ modulo $p$.

Using the same idea, one can also handle tree languages defined by
counting the number of occurrences of certain letters modulo $p$
threshold $q$. It suffices to consider, in analogy with the mod $p$
case, the languages of the form $K_{k}(\exists^r_{p,q})$, and the
preclone $T_{p,q}$, a sub-preclone of $T(\B_{p+q})$, whose rank $n$
elements are the mappings $f_{r}\colon (r_{1},\ldots,r_{n})\mapsto
r_{1}+\cdots +r_{n}+r$, where the sum is taken modulo $p$ threshold
$q$. Note that this notion generalizes both above examples, since
$T_{p} = T_{p,0}$ and that $T_{\exists} = T_{1,1}$.

\paragraph*{Identification of a path}
Let $K_{k}(\path)$ be the set of all the trees in $\Delta M_k$ such
that all the vertices along at least one maximal path from the root to
a leaf are labeled $1_{n}$ (for the appropriate values of $n$). Then
$K_{k}(\path)$ is recognized by the preclone $T_{\path}$ (see
Example~\ref{define Tpath}).

Let indeed $\alpha\colon \Delta M\to T_{\path}$ be the morphism given
by $\alpha(0_{n}) = \false_{n}$, $\alpha(1_{0}) = \true_{0}$ and
$\alpha(1_{n}) = \sfor_{n}$ ($n\ne 0$). One can then verify that
$\alpha\inv(\true_{k}) = K_{k}({\path})$.
	
\paragraph*{Identification of the \next\ modality}
Let $K_{k}(\next)$ consist of all the trees in $\Delta M_k$
such that each maximal path has length at least two and the children of
the root are labeled $1_{n}$ (for the appropriate $n$). We show that
$K_{k}(\next)$ is recognizable.

Recall that $\B = \{\true,\false\}$, and let $\alpha\colon \Delta M \to
T(\B\times\B)$ be the  morphism given as follows:
\begin{itemize}
    \item $\alpha(0_{0})$ is the nullary constant
    $(\false,\false)_{0}$,
    
    \item $\alpha(1_{0})$ is the nullary constant $(\false,\true)_{0}$,
    
    \item if $n> 0$, then $\alpha(0_{n})$ is the $n$-ary map
    $((x_{1},y_{1}),\ldots,(x_{n},y_{n})) \mapsto
    (\land_{i}y_{i},\false)$
   
    \item if $n> 0$, then $\alpha(1_{n})$ is the $n$-ary map
    $((x_{1},y_{1}),\ldots,(x_{n},y_{n})) \mapsto
    (\land_{i}y_{i},\true)$.
\end{itemize}    
One can verify by structural induction that for each element $x\in
\Delta M_{k}$, the second component of $\alpha(x)$ is $\true$ if and
only if the root of $x$ is labeled $1_{n}$ for some $n$, and the first
component of $\alpha(x)$ is $\true$ if and only if every child of the
root of $x$ is labeled $1_{n}$ for some $n$, that is, if and only if
$x\in K_{k}(\next)$. Thus $K_{k}(\next)$ is recognized by the morphism
$\alpha$.

%%%%%%%%%%%%%%%%%%%%%%%%%%%%%%%%%%%%%%%%%%%%%%%%%%%%%%
\section{Pseudovarieties of preclones} \label{sec psv}

In the usual setting of one-sorted algebras, a pseudovariety is a class
of finite algebras closed under taking finite direct products,
sub-algebras and quotients. Because we are dealing with preclones,
which are infinitely sorted, we need to consider finitary algebras
instead of finite ones, and to adopt more constraining closure
properties in the definition. (We discuss in Remark~\ref{cor-relative}
an alternative approach, which consists in introducing stricter
finiteness conditions on the preclones themselves, namely in
considering only finitely generated, finitely determined, finitary
preclones.)

We say that a class of finitary preclones is a \emph{pseudovariety} if
it is closed under finite direct product, sub-preclones, quotients,
finitary unions of $\omega$-chains and finitary inverse limits of
$\omega$-diagrams. Here, we say that a union $T = \bigcup_{n}T^{(n)}$
of an $\omega$-chain of preclones $T^{(n)}$, $n \geq 0$ is finitary
exactly when $T$ is finitary. Finitary inverse limits $\lim_n T^{(n)}$
of $\omega$-diagrams $\phi_n\colon T^{(n+1)} \to T^{(n)}$, $n \geq 0$
are defined in the same way.

\begin{remark}
    To be perfectly rigorous, we actually require pseudovarieties to be
    closed under taking preclones \emph{isomorphic to} a finitary
    $\omega$-union or to a finitary inverse limit of an
    $\omega$-diagram of their elements.
\end{remark}    

\begin{remark}\label{def inverse limit}
    Recall that the inverse limit $T$ of the $\omega$-diagram
    $(\phi_n)_{n\geq 0}$, written $T = \lim_{n}T^{(n)}$ if the
    $\phi_{n}\colon T^{(n+1)}\to T^{(n)}$ are clear, is the
    sub-preclone of the direct product $\prod_n T^{(n)}$ whose set of
    elements of rank $m$ consists of those sequences $(x_n)_{n\geq 0}$
    with $x_n \in T^{(n)}_m$ such that $\varphi_n(x_{n+1}) = x_n$, for
    all $n \geq 0$. We call the coordinate projections $\pi_p: \lim_n
    T^{(n)} \to T^{(p)}$ the \emph{induced projection morphisms}.
    
    \begin{center}
    \begin{picture}(80,22)(0,-22)
%     \put(0,-41){\framebox(106,41){}}
    \gasset{Nw=14.0,Nframe=n}
    \node[NLangle=0.0,Nw=8.0,Nmr=0.0](n0)(20.0,-2.0){$T$}

    \node[NLangle=0.0,Nw=12.0,Nmr=0.0](n2)(20.0,-20.0){$T^{(n+1)}$}

    \node[NLangle=0.0,Nw=8.0,Nmr=0.0](n3)(0.0,-20.0){$\cdots$}

    \node[NLangle=0.0,Nw=8.0,Nmr=0.0](n4)(6.0,-10.0){$\cdots$}

    \node[NLangle=0.0,Nw=10.0,Nmr=0.0](n5)(40.0,-20.0){$T^{(n)}$}

    \node[NLangle=0.0,Nw=8.0,Nmr=0.0](n6)(60.0,-20.0){$\cdots$}

    \node[NLangle=0.0,Nw=10.0,Nmr=0.0](n8)(80.0,-20.0){$T^{(0)}$}

    \node[NLangle=0.0,Nw=8.0,Nmr=0.0](n9)(47.0,-10.0){$\cdots$}

    \drawedge[ELside=r](n0,n2){$\pi_{n+1}$}

    \drawedge(n0,n5){$\pi_n$}

    \drawedge(n3,n2){}

    \drawedge[ELside=r](n2,n5){$\phi_n$}

    \drawedge(n5,n6){}

    \drawedge(n6,n8){}

    \end{picture}
    \end{center}    
    
    The inverse limit has the following universal property. Whenever
    $S$ is a preclone and the morphisms $\psi_n\colon S \to T^{(n)}$
    satisfy $\psi_n = \varphi_n \circ \psi_{n+1}$ for each $n \geq 0$,
    then there is a unique morphism $\psi\colon S \to \lim_n T^{(n)}$
    with $\pi_n \circ \psi = \psi_n$, for all $n$. This morphism $\psi$
    maps an element $s \in S$ to the sequence $(\psi_n(s))_{n\geq 0}$.
\end{remark}

\begin{expl}\label{limit of finitary not finitary}
    Here we show that the inverse limit of an $\omega$-diagram of
    1-gen\-er\-ated finitary preclones needs not be finitary. Let $\Sigma =
    \{\sigma\}$, where $\sigma$ has rank 1 and consider the free
    preclone $\Sigma M$. Note that $\Sigma M$ has only elements of rank
    1, and that $\Sigma M_{1}$ can be identified with the monoid
    $\sigma^*$. For each $n \geq 0$, let $\approx_n$ be the congruence
    defined by letting $\sigma^k \approx_{n} \sigma^\ell$ if and only
    if $k = \ell$, or $k,\ell \geq n$. Let $T^{(n)} = \Sigma
    M/{\approx_n}$. Then $T^{(n)}$ is again $\sigma$-generated, and it
    can be identified with the monoid $\{0,1,\ldots,n\}$ under addition
    threshold $n$. In particular, $T^{(n)}$ is a finitary preclone.
    Since $\approx_{n+1}$-equivalent elements of $\Sigma M$ are also
    $\approx_n$-equivalent, there is a natural morphism of preclones from
    $T^{(n+1)}$ to $T^{(n)}$, mapping $\sigma$ to itself, and the
    inverse limit of the resulting $\omega$-diagram is $\Sigma M$
    itself, which is not finitary.
\end{expl}

Pseudovarieties of preclones can be characterized using the notion of
{\em division}: we say that a preclone $S$ \emph{divides} a preclone
$T$, written $S < T$, if $S$ is a quotient of a sub-preclone of $T$. It
is immediately verified that a nonempty class of finitary preclones is
a pseudovariety if and only if it is closed with respect to division,
binary direct product, finitary unions of $\omega$-chains and finitary
inverse limits of $\omega$-diagrams.

\begin{expl}
    It is immediate that the intersection of a collection of
    pseudovarieties of preclones is a pseudovariety. It follows that if
    \K\ is a class of finitary preclones, then the pseudovariety
    generated by \K\ is well defined, as the least pseudovariety
    containing \K. In particular, the elements of this pseudovariety,
    written $\langle\K\rangle$, can be described in terms of the
    elements of \K, taking subpreclones, quotients, direct products,
    finitary unions of $\omega$-chains and inverse limits of
    $\omega$-diagrams. See Section~\ref{psv generated} below.

    We discuss other examples in Section \ref{sec examples tree
    varieties}.
\end{expl}    

We first explore the relation between pseudovarieties and their
finitely determined elements, then we discuss pseudovarieties generated
by a class of preclones, and finally, we explore some additional
closure properties of pseudovarieties.

% % % % % % % % % % % % % % % % 
\subsection{Pseudovarieties and their finitely determined elements}
\label{sec psv vs fd}

\begin{prop}\label{prop-limit}
    Let $S$ be a preclone.
    \begin{enumerate}
	\item $S$ is isomorphic to the inverse limit $\lim_n S^{(n)}$
	of an $\omega$-diagram, where each $S^{(n)}$ is an
	$n$-determined quotient of $S$.
	
	\item If $S$ is finitary, then $S$ is isomorphic to the union
	of an $\omega$-chain $\bigcup_{n\geq 0} T^{(n)}$, where each
	$T^{(n)}$ is the inverse limit of an $\omega$-diagram of
	finitely generated, finitely determined divisors of $S$.
    \end{enumerate}	
\end{prop} 

\proof  
Let $S^{(n)} = S/{\sim_n}$ (where $\sim_{n}$ is defined in
Section~\ref{sec representation}) and let $\pi_{n}\colon S\rightarrow
S^{(n)}$ be the corresponding projection. Since $\sim_{n+1}$-related
elements of $S$ are also $\sim_{n}$-related, there exists a morphism of
preclones $\phi_n\colon S^{(n+1)} \to S^{(n)}$ such that $\pi_{n} =
\phi_{n}\circ \pi_{n+1}$. Thus the $\pi_{n}$ determine a morphism
$\pi\colon S \to \lim_{n} S^{(n)}$, such that $\pi(s) =
(\pi_{n}(s))_{n}$ for each $s\in S$ (Remark~\ref{def inverse limit}).

Moreover, since $\sim_{n}$ is the identity relation on the elements of
$S$ of rank at most $n$, we find that for each $k\leq n$, $\pi_{n}$
establishes a bijection between the elements of rank $k$ of $S$ and
those of $S^{(n)}$. In particular, $\pi$ is injective since each
element of $S$ has rank $k$ for some finite integer $k$. Furthermore,
for each $k\leq n$, $\phi_{n}$ establishes a bijection between the
elements of rank $k$, and it follows that each element of rank $k$ of
$\lim_{n} S^{(n)}$ is the $\pi$-image of its $k$-th component. That is,
$\pi$ is onto. Finally, Lemma~\ref{simk k determined} shows that each $S^{(n)}$
is $n$-determined. This concludes the proof of the first statement.

We now assume that $S$ is finitary, and we let $T^{(m)}$ be the
sub-preclone generated by the elements of $S$ of rank at most $m$.
Then $T^{(m)}$ is finitely generated, and the first statement shows
that $T^{(m)}$ is the inverse limit of an $\omega$-diagram of
finitely generated, finitely determined quotients of $T^{(m)}$, which
are in particular divisors of $S$.
\eop

The following corollary follows immediately.

\begin{cor}\label{cor-k-faithful}
    Every pseudovariety of preclones is uniquely determined by its
    finitely generated, finitely determined elements.
\end{cor}

We can go a little further, and show that a pseudovariety is
determined by the syntactic preclones it contains.

\begin{prop}\label{subdirect product of syntactic}
    Let $S$ be a finitely generated, $k$-determined, finitary preclone, let $A$
    be a finite ranked set and let $\phi\colon AM \to S$ be an onto
    morphism. Then $S$ divides the direct product of the syntactic
    preclones of the languages $\phi\inv(s)$, where $s$ runs over the
    (finitely many) elements of $S$ of rank at most $k$.
\end{prop}

\proof
It suffices to show that if $x,y\in AM_{n}$ for some $n\ge 0$ and
$x\sim_{\phi\inv(s)} y$ for each $s\in S_{\ell}$, $\ell\leq k$, then
$\phi(x) = \phi(y)$.

First, suppose that $x$ and $y$ have rank $n \leq k$, and let $s =
\phi(x)$. Then $(\1,0,\n,0)$ is a $\phi\inv(s)$-context of $x$, so it
is also a $\phi\inv(s)$-context of $y$, and we have $\phi(y) = s =
\phi(x)$. Now, if $x$ and $y$ have rank $n > k$, let $v\in S_{n,p}$ for
some $p\leq k$. Since $\phi$ is onto, there exists an element $z\in
AM_{n,p}$ such that $\phi(z) = v$. For each $s\in S_{\ell}$, $\ell\leq
k$, we have $x\sim_{\phi\inv(s)} y$, and hence also $x\cdot
z\sim_{\phi\inv(s)} y\cdot z$. The previous discussion shows therefore
that $\phi(x\cdot z) = \phi(y\cdot z)$, that is, $\phi(x)\cdot v =
\phi(y)\cdot v$. Since $S$ is $k$-determined, it follows that $\phi(x)
= \phi(y)$.
\eop

\begin{cor}\label{psv determined by syntactic}
    Every pseudovariety of preclones is uniquely determined by the
    syntactic preclones it contains.
\end{cor}

\proof
This follows directly from Corollary \ref{cor-k-faithful} and
Proposition \ref{subdirect product of syntactic}.
\eop

% % % % % % % % % % % % % % % % 
\subsection{The pseudovariety generated by a class of preclones}
\label{psv generated}

Let $\I,\H,\S,\P,\L,\U$ denote respectively the operators of taking all
isomorphic images, homomorphic images, subpreclones, finite direct
products, finitary inverse limits of an $\omega$-diagram, and finitary
$\omega$-unions over a class of finitary preclones. The following fact
is a special case of a well-known result in universal algebra.

\begin{lem}\label{lem-HSP}
    If $\K$ is a class of finitary preclones, then $\HSP(\K)$ is the
    least class of finitary preclones containing $\K$, closed under
    homomorphic images, subpreclones and finite direct products.
\end{lem}

Next, we observe the following elementary facts.

\begin{lem}\label{elementary psv}
For all classes $\K$ of finitary preclones, we have
\begin{eqnarray*}
    (1)\enspace\P\L(\K) \subseteq \L\P(\K),&\qquad\qquad&
    (2)\enspace\P\U(\K) \subseteq \U\P(\K),\\
    (3)\enspace\S\L(\K) \subseteq \L\S(\K),&\qquad\qquad&
    (4)\enspace\S\U(\K) \subseteq \U\S(\K).
\end{eqnarray*}
\end{lem}

\proof
To prove the first inclusion, suppose that $S$ is the direct product of
the finitary preclones $S^{(i)}$, $i\in [n]$, where each $S^{(i)}$ is a
limit of an $\omega$-diagram of preclones $S^{(i,k)}$ in $\K$
determined by a family of morphisms $\varphi_{i,k}\colon S^{(i,k+1)} \to
S^{(i,k)}$, $k \geq 0$. For each $k$, let $T^{(k)}$ be the direct
product $\prod_{i \in [n]}S^{(i,k)}$, and let $\varphi_k = \prod_{i \in
[n]} \varphi_{i,k}\colon T^{(k+1)} \to T^{(k)}$. It is a routine matter to
verify that $S$ is isomorphic to the limit of the $\omega$-diagram
determined by the family of morphisms $\varphi_k\colon T^{(k+1)} \to
T^{(k)}$, $k \geq 0$. Thus, $S \in \L\P(\K)$.

Now, for each $i \in [n]$, let $(S^{(i,k)})_{k \geq 0}$ be an
$\omega$-chain of finitary preclones in $\K$. Let us assume that each
$S^{(i)} = \bigcup_{k \geq 0}S^{(i,k)}$ is finitary, and let $S =
\prod_{i \in [n]} S^{(i)}$. If $s = (s_1,\ldots,s_n) \in S$, then each
$s_i$ belongs to $S^{(i,k_i)}$, for some $k_i$. Thus $s\in \prod_{i \in
[n]} S^{(i,k)}$, where $k = \max k_i$, and we have shown that $S \in
\bigcup_{k \geq 0} \prod_{i \in [n]}S^{(i,k)}$, so that $S \in
\U\P(\K)$.

To prove the third inclusion, let $T$ be a sub-preclone of $\lim_{n}
S^{(n)}$, the finitary inverse limit of an $\omega$-diagram
$\phi_{n}\colon S^{(n+1)}\to S^{(n)}$ of elements of $\K$. Let
$\pi_{n}\colon T\to S^{(n)}$ be the natural projections (restricted to
$T$), and let $T^{(n)} = \pi_{n}(T)$. Then $T^{(n)}$ is a subpreclone
of $S^{(n)}$ for each $n$. Moreover, the restrictions of the $\phi_{n}$
to $T^{(n+1)}$ define an $\omega$-diagram of subpreclones of elements
of $\K$, and it is an elementary verification that $T = \lim_{n}
T^{(n)}$. Since $T$ is finitary, we have proved that $T \in \L\S(\K)$.

As for the last inclusion, let $T$ be a subpreclone of a finitary union
$\bigcup_{k \geq 0}S^{(k)}$ with $S^{(k)} \in \K$, for all $k\geq 0$.
Let $T^{(k)} = S^{(k)} \cap T$ for each $k \geq 0$. Then each $T^{(k)}$
is a subpreclone of $S^{(k)}$ and $T = \bigcup_{k \geq 0} T^{(k)}$. It
follows that $T \in \U\S(\K)$.
\eop

Our proof of the third inclusion actually yields the following result.

\begin{cor}\label{smart proof}
    If a finitary preclone $S$ embeds in an inverse limit $\lim_n
    S^{(n)}$, then $S$ is isomorphic to a (finitary) inverse limit
    $\lim_n T^{(n)}$, where each $T^{(n)}$ is a finitary sub-preclone of
    $S^{(n)}$.
\end{cor}

We can be more precise than Lemma~\ref{elementary psv} for what
concerns finitely generated, finitely determined preclones.

\begin{lem}\label{union vs finitely determined}
    Let $T$ be a preclone which embeds in the union of an
    $\omega$-chain $(S^{(n)})_{n}$. If $T$ is finitely generated,
    then $T$ embeds in $S^{(n)}$ for all large enough $n$.
\end{lem}

\proof
Since $T$ is finitely generated, its set of generators is
entirely contained in some $S^{(k)}$, and hence $T$ embeds in each
$S^{(n)}$, $n\geq k$.
\eop

\begin{lem}\label{lem-HU}
    Let $T$ be a quotient of the union of an $\omega$-chain
    $(S^{(n)})_{n}$. If $T$ is finitely generated, then $T$ is a
    quotient of $S^{(n)}$ for all large enough $n$.
\end{lem}

\proof
Let $\phi$ be a surjective morphism from $S = \bigcup_{n}S^{(n)}$ onto
$T$. Since $T$ is finitely generated, there exists an integer $k$ such
that $\phi(S^{(k)})$ contains all the generators of $T$, and this
implies that the restriction of $\phi$ to $S^{(k)}$ (and to each
$S^{(n)}$, $n\geq k$) is onto.
\eop 

\begin{lem}\label{invlim vs finitely determined}
    Let $T$ be a preclone which embeds in the inverse limit
    $\lim_{n}S^{(n)}$ of an $\omega$-diagram, and for each $n$, let
    $\pi_{n}\colon T\rightarrow S^{(n)}$ be the natural projection
    (restricted to $T$). If $T$ is finitary, then for each $k$,
    $\pi_{n}$ is $k$-injective for all large enough $n$. If in addition
    $T$ is finitely determined, then $T$ embeds in $S_{n}$ for all
    large enough $n$.
\end{lem}

\proof 
Since $T$ is finitary, $T_{k}$ is finite for each integer $k$, and
hence there exists an integer $n_{k}$ such that $\pi_{n}$ is injective
on $T_{k}$ for each $n\geq n_{k}$. In particular, for each integer $k$,
$\pi_{n}$ is $k$-injective for all large enough $n$. The last part of
the statement follows from Lemma~\ref{prop-k-injective}.
\eop

\begin{lem}\label{lem-HL}
    Let $T$ be a quotient of the finitary inverse limit $\lim_{n}
    S^{(n)}$ of an $\omega$-diagram. If $T$ is finitely determined,
    then $T$ is a quotient of a sub-preclone of one of the $S^{(n)}$.
\end{lem}

\proof
Let $S = \lim_{n}S^{(n)}$ and let $\pi_{n}\colon S\rightarrow S^{(n)}$
be the corresponding projection. Let also $\phi\colon S\rightarrow T$
be an onto morphism, and let $k\ge 0$ be an integer such that $T$ is
$k$-determined. By Lemma~\ref{invlim vs finitely determined}, $\pi_{n}$ is
$k$-injective for some integer $n$.

Consider the preclone $\pi_{n}(S) \subseteq S^{(n)}$. Then we claim
that the assignment $\pi_{n}(s) \mapsto \varphi(s)$ defines a
surjective morphism $\pi_{n}(S) \to T$. The only nontrivial point is
to verify that this assignment is well defined. Let $s,s'\in S_{p}$ and
suppose that $\pi_{n}(s) = \pi_{n}(s')$. We want to show that $\phi(s)
= \phi(s')$, and for that purpose, we show that $\phi(s)\cdot v =
\phi(s')\cdot v$ for each $v\in T_{p,\ell}$, $\ell\leq k$ (since $T$ is
$k$-determined). Since $\phi$ is onto, there exists $w\in S_{p,\ell}$
such that $v = \phi(w)$. In particular, $\phi(s)\cdot v = \phi(s\cdot
w)$ and similarly, $\phi(s')\cdot v = \phi(s'\cdot w)$. Moreover, we
have $\pi_{n}(s\cdot w) = \pi_{n}(s'\cdot w)$. Now $s\cdot w$ and
$s'\cdot w$ lie in $S_{\ell}$, and $\pi_{n}$ is injective on
$S_{\ell}$, so $s\cdot w = s'\cdot w$. It follows that $\phi(s)\cdot v
= \phi(s')\cdot v$, and hence $\phi(s) = \phi(s')$.
\eop 

We are now ready to describe the finitely generated, finitely
determined elements of the pseudovariety generated by a given class of
finitary preclones.

\begin{prop}\label{psv generated by finitely determined}
    Let $\K$ be a class of finitary preclones. A finitely generated,
    finitely determined, finitary preclone belongs to the pseudovariety
    $\langle\K\rangle$ generated by $\K$ if and only if it divides a
    finite direct product of preclones in $\K$, i.e., it lies in
    $\HSP(\K)$.
\end{prop} 

\proof
It is easily verified that $\langle\K\rangle = \bigcup_{n}\bV_{n}$,
where $\bV_{0} = \K$ and $\bV_{n+1} = \HSP\U\HSP\L(\bV_{n})$. We show
by induction on $n$ that if $T$ a finitely generated, finitely
determined preclone in $\bV_{n}$, then $T\in\HSP(\K)$.

The case $n = 0$ is trivial and we now assume that $T\in \bV_{n+1}$. By
Lemma~\ref{elementary psv}, $T$ lies in $\H\U\S\P\H\L\S\P(\bV_{n})$.
Then Lemma~\ref{lem-HU} shows that $T$ is in fact in
$\HSP\H\L\S\P(\bV_n)$, which is equal to $\HSP\L\S\P(\bV_{n})$ by
Lemma~\ref{lem-HSP}, and is contained in $\H\L\S\P(\bV_{n})$ by
Lemma~\ref{elementary psv} again. Now Lemma~\ref{lem-HL} shows that $T$
lies in fact in $\HSP(\bV_n)$, and we conclude by induction that $T\in
\HSP(\K)$.
\eop

\begin{cor}
    If $\K$ is a class of finitary preclones, then $\langle\K\rangle =
    \I\ULHSP(\K)$.
\end{cor}

\proof
The containment $\I\ULHSP(\K) \subseteq \langle\K\rangle$ is immediate.
To show the reverse inclusion, we consider a finitary preclone $T \in
\langle\K\rangle$. Then $T = \bigcup T^{(n)}$, where $T^{(n)}$ denotes
the subpreclone of $T$ generated by the elements of rank at most $n$.
Now each $T^{(n)}$ is finitely generated, and by
Proposition~\ref{prop-limit}, it is isomorphic to the inverse limit of
the $\omega$-diagram formed by the finitely generated, finitely
determined preclones $T_n/{\sim_m}$, $m \geq 0$. By the
Proposition~\ref{psv generated by finitely determined}, each of these
preclones is in $\HSP(\K)$, so $T \in \I\ULHSP(\K)$.
\eop

\begin{remark}\label{cor-relative}
    As indicated in the first paragraph of Section~\ref{sec psv},
    Proposition~\ref{psv generated by finitely determined} hints at an
    alternative treatment of the notion of pseudovarieties of
    preclones, limited to the consideration of finitely generated,
    finitely determined, finitary preclones. Say that a class $\K$ of
    finitely generated, finitely determined, finitary preclones is a
    \emph{relative pseudovariety} if whenever a finitely generated,
    finitely determined, finitary preclone $S$ divides a finite direct
    product of preclones in $\K$, then $S$ is in fact in $\K$. For each
    pseudovariety $\bV$, the class $\bV_{\sf fin}$ of all its finitary,
    finitely generated, finitely determined members is a relative
    pseudovariety, and the map $\bV\mapsto \bV_{\sf fin}$ is injective
    by Corollary~\ref{cor-k-faithful}. Moreover, Proposition~\ref{psv
    generated by finitely determined} can be used to show that this map
    is onto. That is, the map $\bV\mapsto \bV_{\sf fin}$ is an
    order-preserving bijective correspondence (with respect to the
    inclusion order) between pseudovarieties and relative
    pseudovarieties of preclones.
\end{remark}

Proposition~\ref{psv generated by finitely determined} also leads to
the following useful result. Recall that a finitely generated preclone
$S$ is effectively given if we are given a finite generating set $A$ as
transformations of finite arity of a given finite set $Q$, see
Section~\ref{preclone vs tree automata}.

\begin{cor}\label{decide 1-generated psv}
    Let $S$ and $T$ be effectively given, finitely generated, finitely
    determined preclones. Then it is decidable whether $T$ belongs to
    the pseudovariety of preclones generated by $S$.
\end{cor}

\proof
Let $A$ (resp. $B$) be the given set of generators of $S$ (resp. $T$)
and let$\bV$ be the pseudovariety generated by $S$. By
Corollary~\ref{psv generated by finitely determined}, $T \in \bV$ if
and only if $T$ divides a direct power of $S$, say, $T < S^m$. Since
$B$ is finite, almost all the sets $B_{k}$ are empty. We claim that the
exponent $m$ can be bounded by
$$\prod_{B_k \neq \emptyset}|A_k|^{|B_k|}.$$
Indeed, there exists a sub-preclone $S' \subseteq S^m$ and an onto
morphism $S'\to T$. Since $B$ generates $T$, we may assume without loss
of generality that this morphism defines a bijection from a set $A'$ of
generators of $S'$ to $B$, and in particular, we may identify $B_{k}$
with $A'_{k}$, a subset of $A^m_{k}$. Next, one verifies that if $m$ is
greater than the bound in the claim, then there exist $1\leq i < j \leq
m$ such that for all $k$ and $x \in A'_k$, the $i$-th and the $j$-th
components of $x$ are equal --- but this implies that the exponent can
be decreased by $1$.

Thus, it suffices to test whether or not $T$ divides $S^m$, where $m$
is given by the above formula. But as discussed above, this holds if
and only if $A^m$ contains a set $A'$ and a rank preserving bijection
from $A'$ to $B$ which can be extended to a morphism from the
sub-preclone of $S^m$ generated by $A'$ to $T$. By
Proposition~\ref{prop-extension}, and since $S$ and $T$ are effectively
given and $T$ is finitely determined, this can be checked
algorithmically.
\eop 

\subsection{Closure properties of pseudovarieties}

Here we record additional closure properties of pseudovarieties of
preclones.

\begin{lem}\label{lem-inv-embed}
    Let $\bV$ be a pseudovariety of preclones and let $T$ be a finitary
    preclone. If $T$ embeds in the inverse limit of an $\omega$-diagram
    of preclones in $\bV$, then $T\in\bV$.
\end{lem}

\proof
The lemma follows immediately from Corollary~\ref{smart proof}.
\eop

\begin{prop}\label{prop311}
    Let $\bV$ be a pseudovariety of preclones and let $S$ be a finitary
    preclone. If for each $n\geq 0$, there exists a morphism
    $\phi_n\colon S \to S^{(n)}$ such that $S^{(n)}\in \bV$ and
    $\phi_n$ is injective on elements of rank exactly $n$, then $S \in
    \bV$.
\end{prop}

\proof
Without loss of generality we may assume that each $\phi_n$ is
surjective. For each $n \geq 0$, consider the direct product $T^{(n)} =
S^{(0)} \times \cdots \times S^{(n)}$, which is in $\bV$, and let
$\mu_n$ denote the natural projection of $T^{(n+1)}$ onto $T^{(n)}$.
Let also $\psi_n\colon S \to T^{(n)}$ be the target tupling of the
morphisms $\phi_i$, $i \leq n$, let $T$ be the inverse limit $\lim_{n}
T^{(n)}$ determined by the morphisms $\mu_n$, and let $\pi_n\colon T\to
T^{(n)}$ be the corresponding projection morphisms.

Note that each $\psi_n$ is $n$-injective, and equals the composite of
$\psi_{n+1}$ and $\mu_n$. Thus, there exists a (unique) morphism
$\psi\colon S\to T$ such that the composite of $\psi$ and $\pi_n$ is
$\psi_n$ for each $n$. It follows from the $n$-injectivity of each
$\psi_n$, that $\psi$ is injective. Thus, $S$ embeds in the inverse
limit of an $\omega$-diagram of preclones in $\bV$, and we conclude by
Lemma~\ref{lem-inv-embed}.
\eop

% We note the following two easy corollaries of Proposition~\ref{prop311}.
We note the following easy corollary of Proposition~\ref{prop311}.

\begin{cor}\label{cor-sep}
    Let $\bV$ be a pseudovariety of preclones. Let $S$ be a finitary
    preclone such that distinct elements of equal rank can be separated
    by a morphism from $S$ to a preclone in $\bV$. Then $S \in \bV$.
\end{cor}

\proof
For any distinct elements $f,g$ of equal rank $n$, let
$\phi_{f,g}\colon S \to S_{f,g}$ be a morphism such that $S_{f,g} \in
\bV$ and $\phi_{f,g}(f) \neq \phi_{f,g}(g)$. For any integer $n$, let
$\phi_{n}$ be the target tupling of the finite collection of morphisms
$\phi_{f,g}$ with $f,g\in S_{n}$. Then $\phi_{n}$ is injective on $S_n$
and we conclude by Proposition~\ref{prop311}.
\eop

% \begin{cor}
%     A nonempty class of finitary preclones is a pseudovariety if and
%     only if it is closed under quotients and it has the following
%     property: Whenever $S$ is a finitary preclone such that for each
%     $n\geq 0$ there is a morphism from $S$ to a preclone in $\bV$ which
%     is injective on $S_n$, then $S$ belongs to $\bV$.
% \end{cor} 
% 

%%%%%%%%%%%%%%%%%%%%%%%%%%%%%%%%%%%%%%%%%%%%%%%%%%%%%%
\subsection{Pseudovarieties of \pgs}

The formal treatment pseudovarieties of \pgs\ is similar to the above
treatment of pseudovarieties of preclones -- but for the following
remarks.

We define a \emph{pseudovariety of \pgs} to be a class of finitary
\pgs\ closed under finite direct product, sub-\pgs, quotients and
finitary inverse limits of $\omega$-diagrams. Our first remark is that,
in this case, we do not need to mention finitary unions of
$\omega$-chains: indeed, finitary \pgs\ are finitely generated, so the
union of an $\omega$-chain, if it is finitary, amounts to a finite
union.

Next, the notion of inverse limit of $\omega$-diagrams of \pgs\ needs
some clarification. Consider a sequence of morphisms of \pgs, say
$\varphi_n\colon (S^{(n+1)},A^{(n+1)}) \to (S^{(n)},A^{(n)})$. That is,
each $\varphi_n$ is a preclone morphism from $S^{(n+1)}$ to $S^{(n)}$,
which maps $A^{(n+1)}$ into $A^{(n)}$. We can then form the inverse
limit $\lim_n S^{(n)}$ of the $\omega$-diagram determined by the
preclone morphisms $\varphi_n$, and the inverse limit $\lim_n A^{(n)}$
determined by the set mappings $\phi_{n}$. The inverse limit $\lim_n
(S^{(n)},A^{(n)})$ of the $\omega$-diagram determined by the morphisms
of \pgs\ $\varphi_n$ (as determined by the appropriate universal limit,
see Remark~\ref{def inverse limit}) is the $pg$-pair $(S,A)$, where $A
= \lim_{n} A^{(n)}$ and $S$ is the subpreclone of $\lim_n S^{(n)}$
generated by $A$. Recall that this inverse limit is called finitary
exactly when $S$ is finitary and $A$ is finite (see Example~\ref{limit
of finitary not finitary}).

We now establish the close connection between this inverse limit and
the inverse limit of the underlying $\omega$-diagram of preclones, when
the latter is finitary.

\begin{prop}
    Let $\varphi_n\colon (S^{(n+1)},A^{(n+1)}) \to (S^{(n)},A^{(n)})$
    be an $\omega$-diagram of \pgs. Let $S = \lim_n S^{(n)}$ and let
    and $(T,A) = \lim_n (S^{(n)},A^{(n)})$. If $S$ is finitary, then $S
    = T$.
\end{prop}

\proof
We need to show that $A$ generates $S$. Without loss of generality, we
may assume that each $\varphi_n$ maps $A^{(n+1)}$ surjectively onto
$A^{(n)}$, and we denote by $\chi_{n}$ the restriction of $\phi_{n}$ to
$A^{(n+1)}$. By definition, $A$ is the inverse limit of the
$\omega$-diagram given by the $\chi_{n}$, and we denote by
$\rho_{n}\colon A\to A^{(n)}$ the corresponding projection. We also
denote by $\chi_{n}$ and $\rho_{n}$ the extensions of these mappings 
to preclone morphisms $A^{(n+1)}M\to A^{(n)}M$ and $AM\to A^{(n)}M$. 
It is not difficult to verify that $AM$ is the inverse limit of the
$\omega$-diagram given by the $\chi_{n}$, and that the $\rho_{n}$ are
the corresponding projections.

\begin{center}
\begin{picture}(50,22)(0,-22)
%     \put(0,-41){\framebox(106,41){}}
\gasset{Nw=14.0,Nframe=n}
\node[NLangle=0.0,Nw=8.0,Nmr=0.0](n0)(18.0,-2.0){$A$}

\node[NLangle=0.0,Nw=12.0,Nmr=0.0](n2)(18.0,-20.0){$A^{(n+1)}$}

\node[NLangle=0.0,Nw=6.0,Nmr=0.0](n3)(0.0,-20.0){$\cdots$}

\node[NLangle=0.0,Nw=8.0,Nmr=0.0](n4)(4.0,-10.0){$\cdots$}

\node[NLangle=0.0,Nw=10.0,Nmr=0.0](n5)(38.0,-20.0){$A^{(n)}$}

\node[NLangle=0.0,Nw=8.0,Nmr=0.0](n6)(45.0,-20.0){$\cdots$}

% \node[NLangle=0.0,Nw=10.0,Nmr=0.0](n8)(80.0,-20.0){$T^{(0)}$}

\node[NLangle=0.0,Nw=8.0,Nmr=0.0](n9)(41.0,-10.0){$\cdots$}

\drawedge[ELside=r](n0,n2){$\rho_{n+1}$}

\drawedge(n0,n5){$\rho_n$}

\drawedge(n3,n2){}

\drawedge[ELside=r](n2,n5){$\chi_n$}

% \drawedge(n5,n6){}

% \drawedge(n6,n8){}

\end{picture}
\qquad
\begin{picture}(52,22)(0,-22)
%     \put(0,-41){\framebox(106,41){}}
\gasset{Nw=14.0,Nframe=n}
\node[NLangle=0.0,Nw=8.0,Nmr=0.0](n0)(20.0,-2.0){$AM$}

\node[NLangle=0.0,Nw=15.0,Nmr=0.0](n2)(20.0,-20.0){$A^{(n+1)}M$}

\node[NLangle=0.0,Nw=7.0,Nmr=0.0](n3)(0.0,-20.0){$\cdots$}

\node[NLangle=0.0,Nw=8.0,Nmr=0.0](n4)(6.0,-10.0){$\cdots$}

\node[NLangle=0.0,Nw=12.0,Nmr=0.0](n5)(43.0,-20.0){$A^{(n)}M$}

% \node[NLangle=0.0,Nw=8.0,Nmr=0.0](n6)(60.0,-20.0){$\cdots$}

% \node[NLangle=0.0,Nw=10.0,Nmr=0.0](n8)(80.0,-20.0){$T^{(0)}$}

\node[NLangle=0.0,Nw=8.0,Nmr=0.0](n9)(45.0,-10.0){$\cdots$}

\node[NLangle=0.0,Nw=8.0,Nmr=0.0](n10)(52.0,-20.0){$\cdots$}

\drawedge[ELside=r](n0,n2){$\rho_{n+1}$}

\drawedge(n0,n5){$\rho_n$}

\drawedge(n3,n2){}

\drawedge[ELside=r](n2,n5){$\chi_n$}

% \drawedge(n5,n6){}

% \drawedge(n6,n8){}

\end{picture}
\end{center}

Moreover, each $\rho_{k}$ is onto (even from $A$ to $A^{(k)}$). Let
indeed $a_{k}\in A^{(k)}$. Since the $\chi_{n}$ are onto, we can
define by induction a sequence $(a_{n})_{n\geq k}$ such that
$\chi_{n}(a_{n+1}) = a_{n}$ for each $n\geq k$. This sequence can be 
completed with the iterated images of $a_{k}$ by $\chi_{k-1}$,
\dots, $\chi_{0}$ to yield an element of $A$ whose $k$-th projection 
is $a_{k}$.

Since $A^{(n)}$ generates $S^{(n)}$, the morphism $\psi_{n}\colon
A^{(n)}M \to S^{(n)}$ induced by $\id_{A^{(n)}}$ is surjective.
Moreover, the composites $\phi_{n}\circ\psi_{n+1}$ and
$\psi_{n}\circ\chi_{n}$ coincide.

\begin{center}
\begin{picture}(60,52)(0,-52)
%     \put(0,-41){\framebox(106,41){}}
\gasset{Nw=14.0,Nframe=n}
\node[NLangle=0.0,Nw=9.0,Nmr=0.0](n0)(0.0,-2.0){$AM$}

\node[NLangle=0.0,Nw=8.0,Nmr=0.0](n1)(0.0,-52.0){$S$}

\node[NLangle=0.0,Nw=15.0,Nmr=0.0](n2)(30.0,-18.0){$A^{(n+1)}M$}

\node[NLangle=0.0,Nw=12.0,Nmr=0.0](n3)(30.0,-36.0){$S^{(n+1)}$}

\node[NLangle=0.0,Nw=12.0,Nmr=0.0](n4)(60.0,-18.0){$A^{(n)}M$}

\node[NLangle=0.0,Nw=12.0,Nmr=0.0](n5)(60.0,-36.0){$S^{(n)}$}

\drawedge(n0,n4){$\rho_n$}

\drawedge[ELside=r](n0,n2){$\rho_{n+1}$}

\drawedge[ELside=r](n1,n5){$\pi_n$}

\drawedge(n1,n3){$\pi_{n+1}$}

\drawedge[ELside=r](n2,n3){$\psi_{n+1}$}

\drawedge(n4,n5){$\psi_{n}$}

\drawedge[ELside=r](n2,n4){$\chi_n$}

\drawedge(n3,n5){$\phi_n$}

\drawedge[dash={2.0 2.0 2.0 3.0}{0.0},ELside=r](n0,n1){$\tau$}

\end{picture}
\end{center}

It follows that the morphisms $\psi_{n}\circ\rho_{n}\colon AM \to
S^{(n)}$ and $\phi_{n}\circ\psi_{n+1}\circ\rho_{n+1}$ coincide, and
hence there exists a morphism $\tau\colon AM \rightarrow S$ such that
$\pi_{n}\circ\tau = \psi_{n}\circ\rho_{n}$ for each $n$. Since
$\rho_{n}$ and $\psi_{n}$ are onto, it follows that each $\pi_{n}$ is
surjective.

We now use the fact that $S$ is finitary. By Lemma~\ref{invlim vs
finitely determined}, $\pi_{n}$ is $k$-injective for each large enough
$n$. Let now $s\in S_{k}$. We want to show that $s\in \tau(AM)$. Let
$n_{k}$ be such that $\pi_{n}$ is $k$-injective for each $n\geq n_{k}$.
We can choose an element $t_{n_{k}}\in A^{(n_{k})}M$ such that
$\psi_{n_{k}}(t_{n_{k}}) = \pi_{n_{k}}(s)$. Then, by induction, we can
construct a sequence $(t_{n})_{n}$ of elements such that
$\chi_{n}(t_{n+1}) = t_{n}$ for each $n\geq 0$. We need to show that 
$\psi_{n}(t_{n}) = \pi_{n}(s)$ for each $n$.

This equality is immediate for $n\leq n_{k}$, and we assume by induction that 
it holds for some $n\geq n_{k}$. We have
$$\phi_{n}(\psi_{n+1}(t_{n+1})) = \psi_{n}(\chi_{n}(t_{n+1})) =
\psi_{n}(t_{n}) = \pi_{n}(s) = \phi_{n}(\pi_{n+1}(t_{n+1})).$$
Since $\pi_{n}$ and $\pi_{n+1}$ are surjective, since they are
injective on $S_{k}$, and since $\phi_{n}\circ\pi_{n+1} = \pi_{n}$,
we find that $\phi_{n}$ is injective on $S^{(n+1)}_{k}$, and hence
$\psi_{n+1}(t_{n+1}) = \pi_{n+1}(s)$, as expected.

Thus $(t_{n})_{n} \in AM$ and $\tau(t) = s$, which concludes the
proof that $S$ is generated by $A$.
\eop

%%%%%%%%%%%%%%%%%%%%%%%%%%%%%%%%%%%%%%%%%%%%%%%%%%%%%
\section{Varieties of tree languages}

Let $\cV = (\cV_{\Sigma,k})_{\Sigma,k}$ be a collection of nonempty
classes of recognizable tree languages $L \subseteq \Sigma M_k$, where
$\Sigma$ runs over the finite ranked alphabet and $k$ runs over the
nonnegative integers. We call $\cV$ a \emph{variety of tree languages},
or a \emph{tree language variety}, if each $\cV_{\Sigma,k}$ is closed
under the Boolean operations, and $\cV$ is closed under inverse
morphisms between free preclones generated by finite ranked sets, and
under quotients defined as follows. Let $L \subseteq \Sigma M_k$ be a
tree language, let $k_{1}$ and $k_{2}$ be nonnegative integers, $u\in
\Sigma M_{k_{1}+1+k_{2}}$ and $v\in \Sigma M_{n,k}$. Then the
\emph{left quotient} $(u,k_1,k_2)^{-1}L$ and the \emph{right quotient}
$Lv^{-1}$ are defined by
\begin{eqnarray*}
    (u,k_1,k_2)^{-1}L &=& \{t \in \Sigma M_{n} \mid u \cdot(\k_1 \oplus
    t \oplus \k_2) \in L\}\quad\hbox{where $k = k_{1}+n+k_{2}$} \\
    Lv^{-1} &=& \{t\in\Sigma M_{n} \mid t \cdot v \in L\},
\end{eqnarray*}
that is, $(u,k_1,k_2)^{-1}L$ is the set of elements of $\Sigma M_{n}$
for which $(u,k_{1},\n,k_{2})$ is an $L$-context, and $Lv\inv$ is the
set of elements of $\Sigma M_{n}$ for which $(\1,0,v,0)$ is an
$L$-context. Below we will write just $u^{-1}L$ for $(u,k_1,k_2)^{-1}L$
if $k_{1}$ and $k_{2}$ are understood, or play no role.

A \emph{literal variety of tree languages} is defined similarly, but
instead of closure under inverse morphisms between finitely generated
free preclones, we require closure under inverse morphisms between
finitely generated free \pgs. Thus, if $L \subseteq \Sigma M_k$ is in a
literal variety $\cV$ and $\phi\colon \Delta M \to \Sigma M$ is a
preclone morphism with $\Sigma, \Delta$ finite and $\phi(\Delta)
\subseteq \Sigma$, then $\phi^{-1}(L)$ is also in $\cV$.

\note{Discuss, here or in the general introduction, the analogy with 
the monoid case, and to its recent developments in terms of
$C$-varieties.}

%%%%%%%%%%%%%%%%%%%%%%%%%%%%%%%%%%%%%%%%%
\subsection{Varieties of tree languages vs. pseudovarieties of
preclones}\label{sec eilenberg}

The aim of this section is to prove an Eilenberg correspondence between
pseudovarieties of preclones (resp. \pgs), and varieties (resp. literal
varieties) of tree languages. For each pseudovariety $\bV$ of preclones
(resp. \pgs), let $\var(\bV) = (\cV_{\Sigma,k})_{\Sigma,k}$, where
$\cV_{\Sigma,k}$ denotes the class of the tree languages $L
\subseteq\Sigma M_k$ whose syntactic preclone (resp. $pg$-pair) belongs
to $\bV$. It follows from Proposition~\ref{cor-synt morph} that
$\var(\bV)$ consists of all those tree languages that can be recognized
by a preclone (resp. $pg$-pair) in $\bV$.

Conversely, if $\cW$ is a variety (resp. a literal variety) of tree
languages, we let $\psv(\cW)$ be the class of all finitary preclones
(resp. \pgs) that only accept languages in $\cW$, \textit{i.e.},
$\alpha^{-1}(F) \subseteq \Sigma M_k$ belongs to $\cW$, for all
morphisms $\alpha\colon \Sigma M \to S$ (resp. $\alpha\colon(\Sigma M,
\Sigma) \to (S,A)$), $k\ge 0$ and $F \subseteq S_k$.

\begin{thm}\label{thm-Eilenberg}
    The mappings $\var$ and $\psv$ are mutually inverse lattice
    isomorphisms between the lattice of pseudovarieties of preclones
    (resp. \pgs) and the lattice of varieties (resp. literal varieties)
    of tree languages.
\end{thm}

\proof
We only prove the theorem for pseudovarieties of \pgs\ and literal
varieties of tree languages. It is clear that for each pseudovariety
$\bV$ of finitary \pgs, if $\var(\bV) = (\cV_{\Sigma,k})_{\Sigma,k}$,
then each $\cV_{\Sigma,k}$ is closed under complementation and contains
the languages $\emptyset$ and $\Sigma M_k$. The closure of
$\cV_{\Sigma,k}$ under union follows in the standard way from the
closure of $\bV$ under direct product: if $L,L'\subseteq \Sigma M_{k}$
are recognized by morphisms into \pgs\ $(S,A)$ and $(S',A')$ in $\bV$,
then $L\cup L'$ is recognized by a morphism into $(S,A)\times(S',A')$.
Thus $\cV_{\Sigma,k}$ is closed under the Boolean operations.

We now show that $\cV$ is closed under quotients. Let $L \subseteq
\Sigma M_k$ be in $\cV_{\Sigma,k}$, let $\alpha\colon (\Sigma M,
\Sigma) \to (S,A)$ be a morphism recognizing $L$ with $(S,A)\in \bV$
and $L = \alpha^{-1}\alpha(L)$, and let $F = \alpha(L)$. Let
$(u,k_{1},v,k_2)$ be an $n$-ary context, that is, $u \in \Sigma M_{k_1
+ 1 + k_2}$, $v\in \Sigma M_{n,\ell}$ and $k_{1}+\ell+k_{2} = k$. Now
let $F' = \{f\in S_{\ell} \mid \alpha(u)\cdot ({\bf k}_1 \oplus f
\oplus {\bf k}_2) \in F\}$. Then for any $t \in \Sigma M_\ell$,
$\alpha(t) \in F'$ if and only if $\alpha(u) \cdot ({\bf k}_1 \oplus
\alpha(t) \oplus {\bf k}_2) \in F$, if and only if $\alpha(u\cdot ({\bf
k}_1 \oplus t \oplus {\bf k}_2)) \in F$ iff $u\cdot ({\bf k}_1 \oplus t
\oplus {\bf k}_2) \in L$. Thus, $\alpha^{-1}(F') = (u, k_1,
k_2)^{-1}L$, which is therefore in $\cV_{\Sigma,\ell}$. Now let $F'' =
\{f \in M_n : f \cdot \alpha(v) \in L\}$. It follows as above that
$Lv^{-1} = \alpha^{-1}(F'')$ and hence $Lv\inv\in \cV_{\Sigma,n}$.

Before we proceed, let us observe that we just showed the following: if
$L\subseteq \Sigma M_{k}$ is a recognizable tree language, then for
each $n \geq 0$ there are only finitely many distinct sets of the form
$((u,k_1,k_2)^{-1}L)v^{-1}$, where $(u,k_{1},v,k_{2})$ is an $n$-ary
context of $\Sigma M_{k}$.

Next, let $\phi\colon (\Sigma M, \Sigma) \to (\Delta M, \Delta)$ be a
morphism of \pgs\ and $L\subseteq \Delta M_{k}$. If $L$ is recognized
by a morphism $\alpha\colon (\Delta M,\Delta) \to (S,A)$, then
$\phi^{-1}(L)$ is recognized by the composite morphism $\phi \circ
\alpha$, and the closure of $\cV$ by inverse morphisms between free
\pgs\ follows immediately. Thus the mapping $\var$ does associate with
each pseudovariety of \pgs\ a literal variety of tree languages, and it
clearly preserves the inclusion order.

Now consider the mapping $\psv$: we first verify that if $\cW$ is a
literal variety of tree languages, then the class $\psv(\cW)$ is a
pseudovariety. Recall that, if $(S,A) < (T,B)$, then any language
recognized by $(S,A)$ is also recognized by $(T,B)$, so if each
language recognized by $(T,B)$ belongs to $\cW$, then the same holds
for $(S,A)$. Note also that any language recognized by the direct
product $(S,A) \times (T,B)$ is a finite union of intersections of the
form $L\cap M$, where $L$ is recognized by $(S,A)$ and $M$ by $(T,B)$;
thus $\psv(\cW)$ is closed under binary direct products. Finally, if
$(S,A) = \lim_n (S^{(n)},A^{(n)})$ is the finitary inverse limit of an
$\omega$-diagram of finitary \pgs, then Lemma~\ref{invlim vs finitely
determined} shows that the languages recognized by $(S,A)$ are
recognized by almost all of the $(S^{(n)},A^{(n)})$. Thus
$(S,A)\in\psv(\cW)$, which concludes the proof that $\psv(\cW)$ is a
pseudovariety of \pgs.

Let $\cW$ be a literal variety of tree languages, and let $\cV =
\var(\psv(\cW))$. We now show that $\cV = \cW$. Since $\cV$ consists of
all the tree languages recognized by a $pg$-pair in $\bW = \psv(\cW)$,
it is clear that $\cV \subseteq \cW$. Now let $L \in\cW_{\Sigma,k}$,
and let $(M_{L},A_{L})$ be its syntactic $pg$-pair. To prove that
$(M_L,A_L) \in \bW$, it suffices to show that if $\alpha\colon (\Sigma
M, \Sigma) \to (M_L,A_L)$ is a morphism of \pgs\ and $x\in M_{L}$, then
$\alpha^{-1}(x) \in \cW$. Since a morphism of \pgs\ maps generators to
generators, up to renaming and identifying letters (which can be done
by morphisms between free \pgs), we may assume that $\alpha$ is the
syntactic morphism of $L$. Thus $\alpha^{-1}(x)$ is an equivalence
class $[w]$ in the syntactic congruence of $L$, and hence
\begin{eqnarray*}
    \alpha^{-1}(x) &=& \bigcap_{w \in
    ((u,k_1,k_2)^{-1}L)v^{-1}} ((u,k_1,k_2)^{-1}L)v^{-1} \cr
%     &&\hbox{\qquad}
    &\cap& \bigcap_{w \not \in ((u,k_1,k_2)^{-1}L)v^{-1}}
    ((u,k_1,k_2)^{-1}\ol{L})v^{-1}
\end{eqnarray*}  
where $\ol{L}$ denotes the complement of $L$. If $x$ has rank $n$, the
intersections in this formula run over $n$-ary contexts
$(u,k_{1},v,k_{2})$, and as observed above, these intersections are in
fact finite. It follows that that $\alpha\inv(x) \in \cV$. This
concludes the verification that $\cV = \cW$, so $\var\circ\psv$ is
the identity mapping, and in particular $\var$ is surjective and
$\psv$ is injective.

It is clear that both maps $\var$ and $\psv$ preserve the inclusion
order. In order to conclude that they are mutually reciprocal
bijections, it suffices to verify that $\var$ is injective. If $\bV$
and $\bW$ are pseudovarieties such that $\var(\bV) = \var(\bW) = \cV$,
then a tree language is in $\cV$ if and only if its syntactic preclone
is in $\bV$, if and only if its syntactic preclone is in $\bW$. Thus
$\bV$ and $\bW$ contain the same syntactic preclones, and it follows
from Corollary~\ref{psv determined by syntactic} that $\bV = \bW$.
\eop

\begin{remark}
     Three further variety theorems for finite trees exist in the
     literature. They differ from the variety theorem proved above
     since they use a different notion of morphism, quotient, and
     syntactic algebra. The variety theorem in \cite{Almeida,Steinby1}
     is formulated for tree language varieties over some fixed ranked
     alphabet and the morphisms are homomorphisms between finitely
     generated free algebras, whereas the ``general variety theorem''
     of \cite{Steinby2} allows for tree languages over different ranked
     alphabets and a more general notion of morphism, closely related
     to the morphisms of free pg-pairs. On the other hand, the
     morphisms in \cite{Esik} are much more general than those in
     either \cite{Almeida,Steinby1,Steinby2} or the present paper, they
     even include nonlinear tree morphisms that allow for the
     duplication of a variable. Another difference is that the tree
     language varieties in \cite{Almeida,Steinby1,Steinby2} involve
     only left quotients, whereas the one presented here (and the
     varieties of \cite{Esik}) are defined using two sided quotients.
     The notion of syntactic algebra is also different in these papers:
     minimal tree automata in \cite{Almeida,Steinby1}, a variant of
     minimal tree automata in \cite{Steinby2}, minimal clone (or
     Lawvere theory) in \cite{Esik}, and minimal preclone, or pg-pair,
     here. We refer to \cite[Section 14]{Esik} for a more detailed
     comparative discussion.
     
     As noted above, the abundance of variety theorems for finite trees
     is due to the fact that there are several reasonable ways of
     defining morphisms and quotients, and a choice of these notions is
     reflected by the corresponding notion of syntactic algebra. No
     variety theorem is known for the 3-sorted algebras proposed in
     \cite{Wilke}.
\end{remark}

%%%%%%%%%%%%%%%%%%%%%%%%%%%%%%%%%%%%%%%%%
\subsection{Examples of varieties of tree languages}
\label{sec examples tree varieties}

%%%%%%%%%%%%%%%%%%%%%%%%%%%%%%%%%%%%%%%%%
\subsubsection{Small examples}

As a practice example, we describe the variety of tree languages
associated with the pseudovariety $\langle T_{\exists}\rangle$
generated by $T_{\exists}$ (see Section~\ref{simple examples}).

Let $\Sigma$ be a finite ranked alphabet and let $L\subseteq \Sigma
M_{k}$ be a tree language accepted by a preclone in $\langle
T_{\exists}\rangle$. Then the syntactic preclone $S$ of $L$ lies in
$\langle T_{\exists}\rangle$. Recall that a syntactic preclone is
finitely generated and finitely determined: it follows from
Proposition~\ref{psv generated by finitely determined} that $S$ divides
a product of a finite number of copies of $T_{\exists}$. By a
standard argument, $L$ is therefore a (positive) Boolean combination 
of languages recognized by a morphism from $\Sigma M$ to
$T_{\exists}$.

Now let $\tau\colon\Sigma M\to T_{\exists}$ be a morphism. As discussed
in Section~\ref{sec more examples trees}, a tree language in $\Sigma M$
recognized by $\tau$ is either of the form $K_{k}(\Sigma')$ for some
$\Sigma'\subseteq \Sigma$, or it is the complement of such a language.
From there, and using the same reasoning as in the analogous case
concerning word languages, one can verify that a language $L\in \Sigma
M_{k}$ is accepted by a preclone in $\langle T_{\exists}\rangle$ if and
only if $L$ is a Boolean combination of languages of the form
$K_{k}(\Sigma')$ ($\Sigma'\subseteq \Sigma$), or equivalently, $L$ is
a Boolean combination of languages of the form $L_{k}(\Sigma')$,
$\Sigma'\subseteq\Sigma$, where $L_{k}(\Sigma)$ is the set of all
$\Sigma$-trees of rank $k$, for which the set of node labels is exactly
$\Sigma'$.

Similarly -- and referring again to Section~\ref{sec more examples
trees} for notation -- one can give a description of the variety of
tree languages associated with the pseudovariety $\langle
T_{p}\rangle$, or the pseudovariety $\langle T_{p,q}\rangle$, using the
languages of the form $K_{k}(\exists_{p}^r)$ or
$K_{k}(\exists_{p,q}^r)$ instead of the $K_{k}(\exists)$.

%%%%%%%%%%%%%%%%%%%%%%%%%%%%%%%%%%%%%%%%%
\subsubsection{$FO[\Suc]$-definable tree languages}\label{sec FOSuc}

In a recent paper \cite{BenediktSegoufin}, Benedikt and S\'egoufin
considered the class of $FO[\Suc]$-definable tree languages. Note that
the logical language used in $FO[\Suc]$ does not allow the predicate
$<$, and $FO[\Suc]$ is a fragment of $FO[<]$. We refer the reader to
\cite{BenediktSegoufin} for precise definitions, and we point out here
that the characterization established there can be expressed in the
framework developed in the present paper.

More precisely, the results of Benedikt and S\'egoufin establish that
$FO[\Suc]$-definable tree languages form a variety of languages, and 
that the corresponding pseudovariety of preclones consists of the
preclones $S$ such that

\begin{itemize}
    \item[(1)] the semigroup $S_{1}$ satisfies $x^\ell = x^{\ell+1}$ and
    $exfyezf = ezfyexf$ for all elements $e,f,x,y,z$ such that $e =
    e^2$ and $f = f^2$ and for $\ell = |S_{1}|$;
    
    \item[(2)] for each $x\in S_{2}$, $e\in S_{1}$ such that $e = e^2$,
    and $s,t\in S_{0}$, $x\cdot(e\cdot s \oplus e\cdot t) =
    x\cdot(e\cdot t \oplus e\cdot s)$.
\end{itemize}    
In particular, $FO[\Suc]$-definability is decidable for regular tree 
languages.

It is clearly argued in \cite{BenediktSegoufin} that
$FO[\Suc]$-definable tree languages are exactly the locally threshold
testable languages, for general model-theoretic reasons, but that this
fact alone does not directly yield a decision procedure. The result
stated above is analogous to the characterization of
$FO[\Suc]$-definability for recognizable word languages - more
precisely, Condition (1) suffices for languages of words and their
syntactic semigroups. Condition (2), which makes sense in trees but not
in words, must be added to the other one to characterize
$FO[\Suc]$-definability for tree languages.

%%%%%%%%%%%%%%%%%%%%%%%%%%%%%%%%%%%%%%%%%
\subsubsection{Some classes of languages definable in modal
logic}\label{sec EFEX}

Boja\'nczyk and Walukiewicz also characterized interesting logically
defined clas\-ses of tree languages \cite{BW}. Again, their results are
not couched in terms of preclones, but they can conveniently be
expressed in this way.

These authors consider three fragments of $\hbox{CTL}^*$: $TL(\EX)$, $TL(\EF)$
and $TL(\EX+\EF)$. Here $\EX$ (resp. $\EF$) denotes the modality
whereby a tree $t$ satisfies $\EX\phi$ (resp. $\EF\phi$) if some child
of the root (resp. some node properly below the root) of $t$ satisfies
$\phi$. The set of formulas constructed using one or both of these
modalities, plus Boolean operations and letter constants form the
logical languages $TL(\EX)$, $TL(\EF)$ and $TL(\EX+\EF)$.

Boja\'nczyk and Walukiewicz first observe that a tree language $L$ is
$TL(\EX)$-definable if and only if there exists an integer $k$ such
that membership of a tree $t$ in $L$ depends only on the fragment of
$t$ consisting of the nodes of depth at most $k$. They then show that
these tree languages form a variety, and the corresponding
pseudovariety of preclones consists of the preclones $S$ such that the
semigroup $S_{1}$ satisfies $ex = e$ for each idempotent $e$. Note that
this is exactly the same characterization as for languages of finite
words \cite{Pin}.

For the characterization of $TL(\EF)$-definable languages, let us first
define the following relation on a preclone $S$ : if $s,t\in S_{n}$, we
say that $s\preceq t$ if $s = u\cdot t$ for some $u\in S_{1}$. It is
easily verified that $\preceq$ is a quasi-order. (The direction of the
order is reversed from that used by Boja\'nczyk and Walukiewicz, to
enhance the analogy with the $\mathcal{R}$- and $\mathcal{L}$-orders in
semigroup theory).

Now let $(S,A)$ be the syntactic $pg$-pair of a tree language
$L\subseteq \Sigma M_{0}$. Then $L$ is $TL(\EF)$-definable if and only
if
\begin{itemize}
    \item $S_{1}$ satisfies the pseudo-identity $v(uv)^\omega =
    (uv)^\omega$ (where $x^\omega$ designates the unique idempotent
    which is a power of $x$); that is, $S_{1}$ is
    $\mathcal{L}$-trivial, and equivalently, the relation $\preceq$ is
    an order relation;
    
    \item $a\cdot(s_{1}\oplus\cdots\oplus s_{n}) =
    a\cdot(s_{\pi(1)}\oplus\cdots\oplus s_{\pi(n)})$ for each $a\in
    A_{n}$ and $s_{1},\ldots, s_{n}\in S_{0}$, and for each permutation
    $\pi$ of $[n]$;
    
    \item $a\cdot(s_{1}\oplus s_{2}\oplus s_{3}\oplus\cdots\oplus
    s_{n}) = a\cdot(s_{2}\oplus s_{2}\oplus s_{3}\oplus\cdots\oplus
    s_{n})$ for each $a\in A_{n}$ and $s_{1},\ldots, s_{n}\in S_{0}$
    such that $s_{2}\preceq s_{1}$;
    
    \item if $b,c\in A_{p}$ and $y \in S_{p,0}$ are such that, for each
    $d\in A_{p}$, we have $d\cdot(b\cdot y\oplus\cdots\oplus b\cdot y)
    = d\cdot y = d\cdot(c\cdot y\oplus\cdots\oplus c\cdot y)$, then
    $a\cdot(z\oplus b\cdot y) = a\cdot(z\oplus c\cdot y)$ for each
    $a\in A_{n}$ and $z\in S_{n-1,0}$.
\end{itemize}
This characterization directly implies the decidability of
$TL(\EF)$-definability.

Boja\'nczyk and Walukiewicz also give an interesting characterization
of the $TL(\EF)$-definable languages in terms of so-called \textit{type
dependency}. In particular, they show that a tree language is
$TL(\EF)$-definable if and only if its syntactic preclone $S$ is such
that, whenever $a$ is the syntactic equivalence class of a letter in
$\Sigma_n$, and the $t_i$'s are syntactic equivalence classes of trees
in $\Sigma M_0$, then the value of a product
$a\cdot(t_{1}\oplus\cdots\oplus t_{n})$ depends only on $a$ and on the
set $\{t \mid t_{i}\preceq t\hbox{ for some }1\le i\le n\}$.

The characterization of $TL(\EX+\EF)$-definable languages given in
\cite{BW} can also be restated in similar -- albeit
more complex -- terms.

%%%%%%%%%%%%%%%%%%%%%%%%%%%%%%%%%%%%%%%%%
\subsubsection{$FO[<]$-definable tree languages}\label{sec FOinf}

The characterization and decidability of $FO[<]$-definable regular tree
languages is an open problem that has attracted some efforts along the
years, as discussed in the introduction.

We obtained an algebraic characterization of $FO[<]$-definable regular
tree languages in terms of pseudovarieties of preclones, as is reported
in \cite{FSTTCS}. A detailed report of this result will appear in
\cite{EWpreparation}, and the present paper lays the foundations for
this proof.

Let us note here that this characterization is analogous to the
characterization of $FO[<]$-definable languages of finite words in the
following sense: it is established in \cite{EWpreparation} that
$FO[<]$-definable tree languages form a variety of tree languages,
whose associated pseudovariety of preclones is the least pseudovariety
containing the preclone $T_{\exists}$ and closed under a suitable
notion of block product. It was pointed out in Example~\ref{define
Texists} that the rank 1 elements of $T_{\exists}$ form the 2-element
monoid $U_{1} = \{1,0\}$, and it is a classical result of language
theory that the least pseudovariety of monoids containing $U_{1}$ and
closed under block product is associated with the variety of
$FO[<]$-definable word languages \cite{Straubing}.

It is also known that, in the word case, this pseudovariety is exactly
that of aperiodic monoids, and membership in it is decidable, which
shows that $FO[<]$-definability is decidable for recognizable word
languages. At the moment, we do not have an analogue of this result,
and we do not know whether $FO[<]$-definability is decidable for
regular tree languages.

Our result \cite{FSTTCS,EWpreparation} actually applies to a larger
class of logically defined regular tree languages, based on the use
of Lindstr\"om quantifiers. First-order logic is thus a particular
case of our result, which also yields (for instance) an algebraic
characterization of first-order logic with modular quantifiers added.

%%%%%%%%%%%%%%%

\thebibliography{99}

{\small

\bibitem{Almeida}
J. Almeida,
On pseudovarieties, varieties of languages,
filters of congruences, pseudoidentities
and related topics,
{\em Algebra Universalis}, 27 (1990), 333--350.

\bibitem{Arnoldetal}
A. Arnold, M.  Dauchet,
Th\'eorie des magmo\"\i des. I. and II.  (in French),
{\em RAIRO Theoretical Informatics and Applications},
12 (1978), 235--257, 3 (1979), 135--154.

\bibitem{BenediktSegoufin}
M. Benedikt, L. S\'egoufin,
Regular tree languages definable in $FO$.
In: \textit{STACS 2005} (V. Diekert, B. Durand eds.), Lect. Notes in
Computer Science, Springer, to appear.

\bibitem{BEbook}
S. L. Bloom, Z. \'Esik,
{\em Iteration Theories}, Springer, 1993.

\bibitem{BW}
M. Boja\'nczyk, I. Walukiewicz,
Characterizing \EF\ and \EX\ tree logics,
in \textit{CONCUR 2004} (P. Gardner, N. Yoshida, eds.), Lect. Notes
in Computer Science 3170, Springer, 2004.

\bibitem{Buchi}
J. R. B\"uchi,
Weak second-order arithmetic and finite automata,
{\em Z. Math. Logik Grundlagen Math.}, 6 (1960), 66--92.

\bibitem{Cohenetal}
J. Cohen, J.-E. Pin, D. Perrin,
On the expressive power of temporal logic,
{\em J. Computer and System Sciences},
46 (1993), 271--294.

\bibitem{tata}
H. Comon, M. Dauchet, R. Gilleron, F. Jacquemard, D. Lugiez, S. Tison,
M. Tommasi, \textit{Tree Automata Techniques and Applications},
available on: \url{http://www.grappa.univ-lille3.fr/tata} (release
October 2002).

\bibitem{Courcelle}
B. Courcelle,
The monadic second-order logic of graphs. I.
Recognizable sets of finite graphs,
{\em Information and Computation}, 85 (1990), 12--75.

\bibitem{Courcelle1996}
B. Courcelle, Basic notions of universal algebra for language theory
and graph grammars, \textit{Theoret. Computer Science}, 163 (1996),
1--54.

\bibitem{CourcelleHdbook}
B. Courcelle, The expression of graph properties and graph
transformations in monadic second order logic. In: G. Rozenberg (ed.)
\textit{Handbook of Graph Grammars and Computing by Graph
Transformations}, vol. 1,  World Scientific, 1997, 313--400.

\bibitem{CourcelleWeil}
B. Courcelle, P. Weil,
The recognizability of sets of graphs is a robust property,
to appear.

\bibitem{Deneckeetal}
K. Denecke, S. L. Wismath,
{\em Universal Algebra and Applications
in Theoretical Computer Science},
Chapman and Hall, 2002.

\bibitem{Diekert}
V. Diekert, \textit{Combinatorics on Traces}, Lect. Notes in Computer Science
454, Springer, 1987.

\bibitem{Doner}
J. Doner,
Tree acceptors and some of their applications,
{\em J. Comput. System Sci.}, 4 (1970), 406--451.

% \bibitem{EbbinghausFlum}
% H.-D. Ebbinghaus and J. Flum,
% {\em Finite Model Theory}, Springer, 1995. 

\bibitem{Eilenberg}
S. Eilenberg,
{\em Automata, Languages, and Machines},
vol. A and B, Academic Press, 1974 and 1976.

\bibitem{EilenbergWright}
S. Eilenberg, J. B.  Wright,
Automata in general algebras,
{\em Information and Control}, 11 (1967), 452--470.

\bibitem{Elgot}
C. C. Elgot,
Decision problems of finite automata design and related arithmetics,
{\em Trans. Amer. Math. Soc.},  98 (1961), 21--51.

\bibitem{Esik}
Z. \'Esik,
A variety theorem for trees and theories,
{\em Publicationes Mathematicae}, 54 (1999), 711--762.

% \bibitem{EsikLarsen}
% Z. \'Esik and K. G. Larsen,
% Regular languages definable by Lindstr\"om quantifiers, 
% {\em Theoretical Informatics and Applications}, to appear.

\bibitem{FSTTCS}
Z. \'Esik, P. Weil,
On logically defined recognizable tree languages.
In: \textit{Proc. FST TCS 2003} (P. K.
Pandya, J. Radhakrishnan eds.), Lect. Notes in Computer
Science  2914 (2003), Springer, 195-207.

\bibitem{EWpreparation}
Z. \'Esik, P. Weil,
Algebraic characterization of logically defined tree languages,
in preparation.

\bibitem{Gabbayetal}
D. M. Gabbay, A. Pnueli, S. Shelah, J. Stavi,
On the temporal analysis of fairness.
In: proc. {\em 12th ACM Symp. Principles of Programming Languages},
Las Vegas, 1980, 163--173.

\bibitem{GecsegSteinby}
F. G\'ecseg, M. Steiby,
{\em Tree Automata},
Akad\'emiai Kiad\'o, Budapest, 1984. 

\bibitem{GSHB}
F. G\'ecseg, M. Steiby, Tree languages. In: G. Rozenberg, A. Salomaa,
eds. \textit{Handbook of Formal Languages}, vol. 3, Springer, 1997,
1-68.

\bibitem{Graetzer}
G. Gr\"atzer, {\em Universal Algebra}, Springer, 1979.

\bibitem{Heuter}
U. Heuter,
First-order properties of trees,
star-free expressions, and aperiodicity. In:
{\em STACS 88} (eds. R. Cori, M. Wirsing), Lect. Notes in Computer
Science 294, Springer, 1988, 136--148.

\bibitem{Kamp}
J. A. Kamp,
Tense logic and the theory of linear order,
Ph. D. Thesis, UCLA, 1968.

% \bibitem{Lindstrom}
% P. Lindstr\"om,
% First order predicate logic with generalized quantifiers.
% {\em Theoria}, 32(1966), 186--195.

\bibitem{MacLane}
S. MacLane, {\em Categories for the Working Mathematician},
Springer, 1971.

\bibitem{McNaughtonPapert}
R. McNaughton, S. Papert,
{\em Counter-Free Automata},
MIT Press, 1971.

\bibitem{MezeiWright}
J. Mezei, J. B. Wright,
Algebraic automata and context-free sets,
{\em Information and Control}, 11 (1967), 3--29.

\bibitem{Pin}
J.-E. Pin,
\textsl{Vari\'et\'es de langages
formels}, Masson, Paris (1984); English translation: \textsl{Varieties
of formal languages}, Plenum, New-York (1986).

\bibitem{Potthoff1}
A. Potthoff, Modulo counting quantifiers over finite trees. In: {\em
CAAP '92} (J.-C. Raoult ed.), Lect. Notes in Computer Science 581,
Springer, 1992, 265--278.

\bibitem{Potthoff95}
A. Potthoff, First order logic on finite trees. In: {\em TAPSOFT '95}
(P.D. Mosses, M. Nielsen, M.I. Schwartzbach eds.), Lect. Notes in
Computer Science 915, Springer, 1995, 125--139.

% \bibitem{RhodesTilson}
% J. Rhodes and B. Tilson,
% The kernel of monoid morphisms,
% {\em J. Pure and Appl. Alg.}, 62(1989),
% 227--268.

\bibitem{Schutzenberger}
M. P. Sch\"utzenberger,  On finite monoids
having only trivial subgroups.
{\em Information and Control}, 8 (1965), 190--194.

\bibitem{Steinby1}
M. Steinby, A theory of tree language varieties. In: {\em Tree automata
and languages} (M. Nivat, A. Podelski, eds.), North Holland, 1992,
57--81.

\bibitem{Steinby2}
M. Steinby,
General varieties of tree languages.
{\em Theoret. Comput. Sci.}, 205 (1998), 1--43.

\bibitem{Straubing}
H. Straubing, {\em Finite Automata, Formal Logic, and Circuit
Complexity,} Birk\-ha\"user, Boston, MA, 1994.

% \bibitem{Straubingetal}
% H. Straubing, D. Therien and W. Thomas,
% Regular languages defined with generalized quantifiers,
% {\em Information and Computation}, 118(1995), 289--301.

\bibitem{ThatcherWright}
J. W. Thatcher, J. B. Wright,
Generalized finite automata theory with
an application to a decision problem of second-order logic,
Math. Systems Theory,  2 (1968), 57--81.

\bibitem{Wechler}
W. Wechler,
\textit{Universal algebra}, EATCS Monographs on Theoretical Computer 
Science 10, Springer, 1992.

\bibitem{WeilMFCS}
P. Weil,
Algebraic recognizability of languages. In:
\it MFCS 2004\rm\ (J. Fiala, V. Koubek, J. Kratochv\'\i l eds.),
Lect. Notes in Computer Science  3153 (Springer, 2004) 149--175.

\bibitem{Wilke}
Th. Wilke, An algebraic characterization of frontier testable tree
languages, {\em Theoret. Comput. Sci.}, 154 (1996), 85--106.

}

%%%%%%%%%%%%%%%%%%%%%%%%%%%%%%%%%%%%%%%%%%%%%%%%%%%%%%
%%%%%%%%%%%%%%%%%%%%%%%%%%%%%%%%%%%%%%%%%%%%%%%%%%%%%%
%%%%%%%%%%%%%%%%%%%%%%%%%%%%%%%%%%%%%%%%%%%%%%%%%%%%%%
%%%%%%%%%%%%%%%%%%%%%%%%%%%%%%%%%%%%%%%%%%%%%%%%%%%%%%
%%%%%%%%%%%%%%%%%%%%%%%%%%%%%%%%%%%%%%%%%%%%%%%%%%%%%%

\end{document}